\documentclass[aps,prx,reprint,amsmath,amssymb,superscriptaddress,nobibnotes,showpacs]{revtex4-2}
\usepackage{kotex}
\usepackage{graphicx}
\usepackage{color}
\usepackage[normalem]{ulem}
\usepackage{amsmath}
\usepackage{mathtools}
\usepackage{natbib}
\usepackage{hyperref}
\usepackage{float}
\usepackage{multirow}
\usepackage{array}
\usepackage{xcolor}	
\usepackage{color, colortbl}
\usepackage{booktabs}
\hypersetup{
    colorlinks=true,
    linkcolor=blue,
    citecolor = blue,
    filecolor=magenta,      
    urlcolor=blue,
    pdfpagemode=FullScreen,
    }

\begin{document}

\title{Distributed quantum machine learning via classical communication}

\author{Kiwmann Hwang}
\email{kiwmann.hwang@gmail.com}
\affiliation{Center for Quantum Technology, Korea Institute of Science and Technology (KIST), Seoul 02792, Korea}
\affiliation{Department of Physics, Yonsei University, Seoul 03722, Korea}
\affiliation{Department of Physics, Korea University, Seoul 02841, Korea}
\author{Hyang-Tag Lim}
\affiliation{Center for Quantum Technology, Korea Institute of Science and Technology (KIST), Seoul 02792, Korea}
\affiliation{Division of Quantum Information, KIST School, Korea University of Science and Technology, Seoul 02792, Korea}
\author{Yong-Su Kim}
\affiliation{Center for Quantum Technology, Korea Institute of Science and Technology (KIST), Seoul 02792, Korea}
\affiliation{Division of Quantum Information, KIST School, Korea University of Science and Technology, Seoul 02792, Korea}
\author{Daniel K. Park}
\affiliation{Department of Applied Statistics, Yonsei University, Seoul 03722, Korea}
\affiliation{Department of Statistics and Data Science, Yonsei University, Seoul 03722, Korea}
\author{Yosep Kim}
\email{yosep9201@gmail.com}
\affiliation{Department of Physics, Korea University, Seoul 02841, Korea}

\date{\today}
\begin{abstract}
Quantum machine learning is emerging as a promising application of quantum computing due to its distinct way of encoding and processing data. It is believed that large-scale quantum machine learning demonstrates substantial advantages over classical counterparts, but a reliable scale-up is hindered by the fragile nature of quantum systems. Here we present an experimentally accessible distributed quantum machine learning scheme that integrates quantum processor units via classical communication. As a demonstration, we perform data classification tasks on 8-dimensional synthetic datasets by emulating two 4-qubit processors and employing quantum convolutional neural networks. Our results indicate that incorporating classical communication notably improves classification accuracy compared to schemes without communication. Furthermore, at the tested circuit depths, we observe that the accuracy with classical communication is no less than that achieved with quantum communication. Our work provides a practical path to demonstrating large-scale quantum machine learning on intermediate-scale quantum processors by leveraging classical communication that can be implemented through currently available mid-circuit measurements.
\end{abstract} 
\maketitle

Quantum superposition and entanglement have revolutionized the paradigm of data encoding and processing, with prototypical quantum algorithms exhibiting computational advantages over classical counterparts~\cite{shor1999polynomial,grover1996fast}.
While a clear computational advantage of quantum processors has not yet been experimentally demonstrated, it is strongly believed that this will be realized along with the development of reliable large-scale quantum processors~\cite{arute2019quantum,pan2022solving,kim2023evidence,tindall2024efficient}. 
Scaling quantum processors involves various strategies, such as increasing a processor's qubit density or integrating multiple processors using distributed architectures~\cite{kimble2008quantum,monroe2014large,bravyi2022future}.
Recent studies underscore that distributed quantum computing is advantageous not only for addressing larger-scale computational problems~\cite{peng2020simulating,eddins2022doubling,piveteau2023circuit} but also for mitigating errors~\cite{katabarwa2024early,avron2021quantum,Lee_2023}.
Accordingly, there is a growing emphasis on adapting known quantum algorithms into distributed forms, including Grover's algorithm~\cite{avron2021quantum,qiu2022distributed}, phase estimation~\cite{li2017application}, and Shor's algorithm~\cite{yimsiriwattana2004distributed,xiao2023distributed}.

\begin{figure}[b!]
\centering
\includegraphics[width=\linewidth]{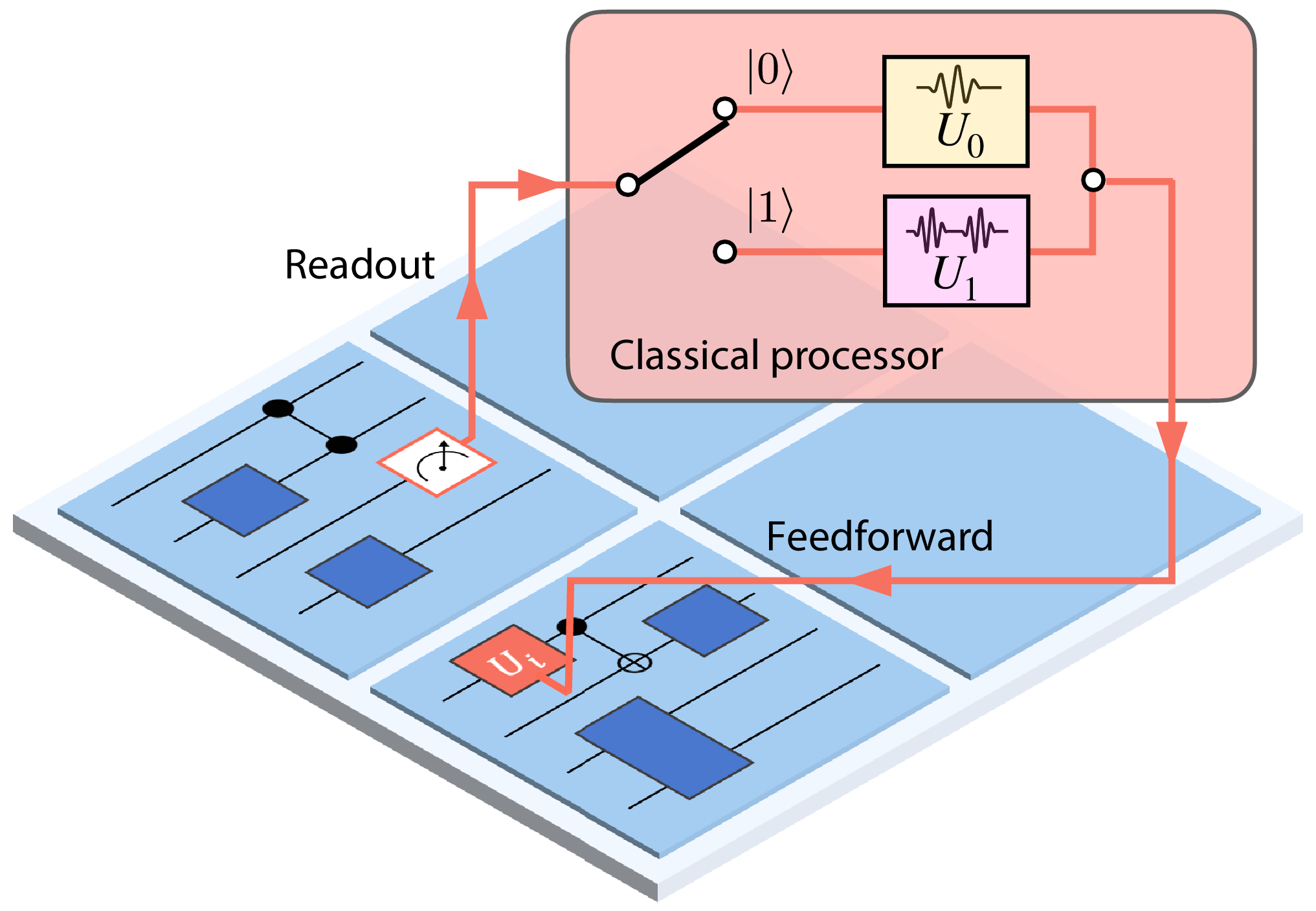}
\caption{\textbf{Schematic of classical communication.} Quantum processor units communicate using mid-circuit measurements and fast feedforward operations. The measurement outcome from one processor determines whether to apply gate $U_0$ or $U_1$ to another processor.
}
\label{fig1}
\end{figure}

\begin{figure*}
\centering
\includegraphics[width=\linewidth]{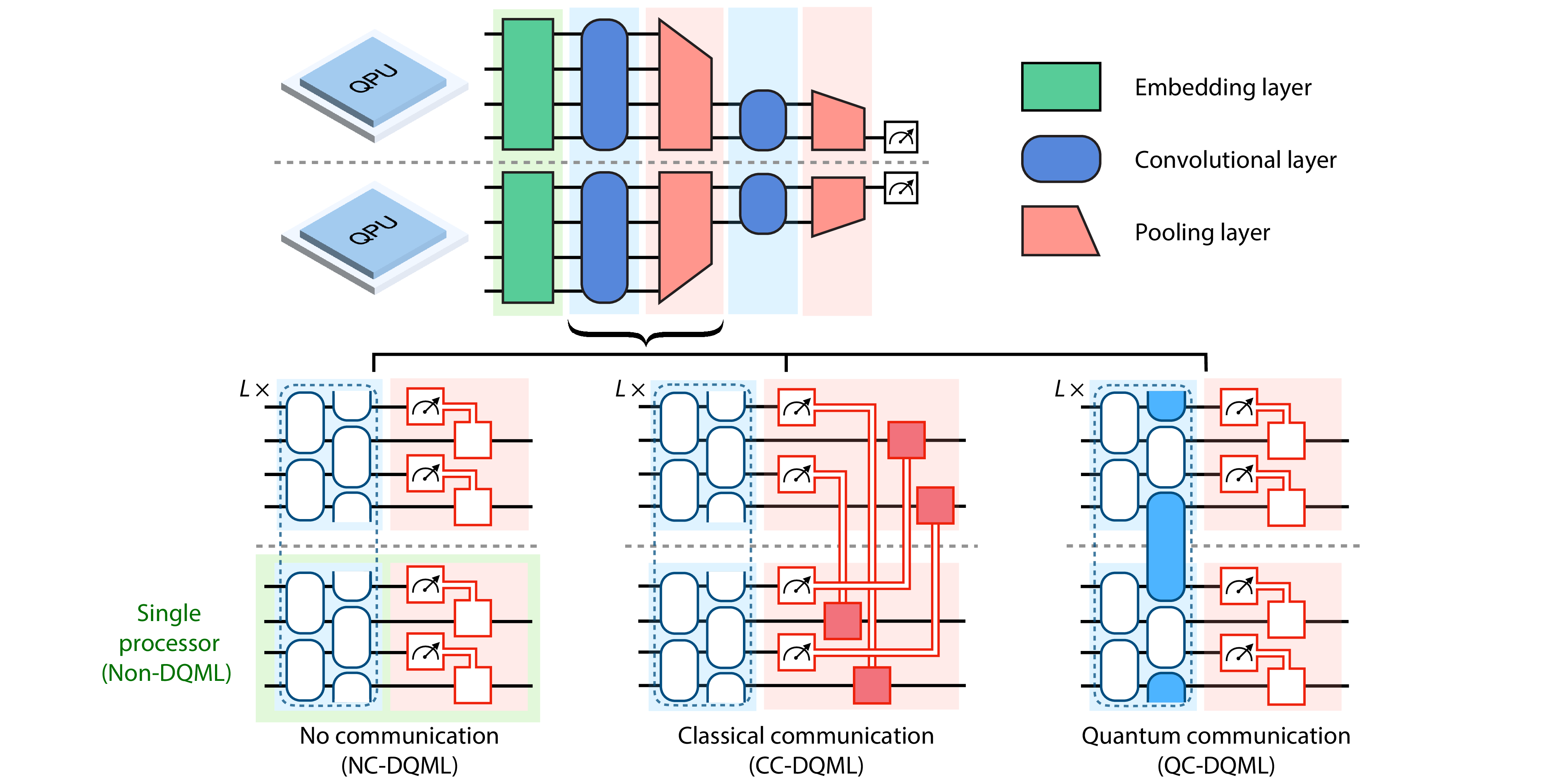}
\caption{\textbf{Distributed quantum machine learning (DQML) models.} The architecture includes an embedding layer that maps data into a Hilbert space, convolutional layers that intertwine data within each quantum processor unit (QPU), and pooling layers that reduce dimensionality. The DQML schemes considered in this work are illustrated at the bottom, with colored operations highlighting the differences. The convolutional layer consists of brick-wall structured convolutional sub-layers, which are repeated to increase the number of training parameters. Classical communication is implemented through feedforward operations
between QPUs, while quantum communication is facilitated by nonlocal two-qubit operations within the convolutional sub-layers. }
\label{fig2}
\end{figure*}

Quantum machine learning, which takes advantage of  quantum contextuality~\cite{deng2017quantum, bowles2023contextuality} and the exponentially large dimensionality of quantum systems~\cite{biamonte2017quantum,havlivcek2019supervised,cerezo2022challenges, abbas2021power}, is no exception. Several practical distributed quantum machine learning (DQML) schemes have been proposed to boost the capabilities of noisy intermediate-scale quantum processors. These schemes are delivering benefits across various domains: processing larger-scale data~\cite{pira2023invitation,marshall2023high}, accelerating training convergence~\cite{du2021accelerating, chen2021federated}, and ensuring data privacy among multiple parties~\cite{neumann2022distributed}.

Most existing DQML schemes rely on executing tasks separately on a single quantum processor~\cite{pira2023invitation,du2021accelerating,chen2021federated,marshall2023high}. While this approach does provide linear increases in computational capacity, greater advantages would be realized by  interconnecting multiple processors through quantum or classical communication. Quantum communication, for example, enables universal quantum operations between separate processors~\cite{gottesman1999demonstrating,eisert2000optimal}. However, achieving high-fidelity entanglement distribution remains a challenge and is currently limited to proof-of-principle experiments~\cite{chou2018deterministic,cacciapuoti2019quantum,zhong2021deterministic}. 
In contrast, classical communication can be reliably implemented within the coherence time, making it more feasible across various physical platforms~\cite{lis2023Midcircuit,corcoles2021exploiting}. While classical communication may offer less advantages than quantum communication~\cite{eddins2022doubling}, it has been shown to exponentially reduce overhead in certain applications compared to schemes without communication~\cite{piveteau2023circuit}. 

In this work, we propose a practical and effective approach for implementing DQML by leveraging classical communication. Our circuit capacity analysis confirms that at shallow depths, our classical communication scheme can produce a range of quantum operations with expressiveness comparable to that of a quantum communication scheme. Furthermore, we observe that, for binary classification tasks on 8-dimensional synthetic datasets, our classical communication scheme significantly outperforms a no-communication scheme and even closely competes with a quantum communication scheme at the tested circuit depths. These results highlight the efficacy of classical communication in expanding the capabilities of small quantum processors.\\

\begin{figure*}
\centering
\includegraphics[width=0.99\linewidth]{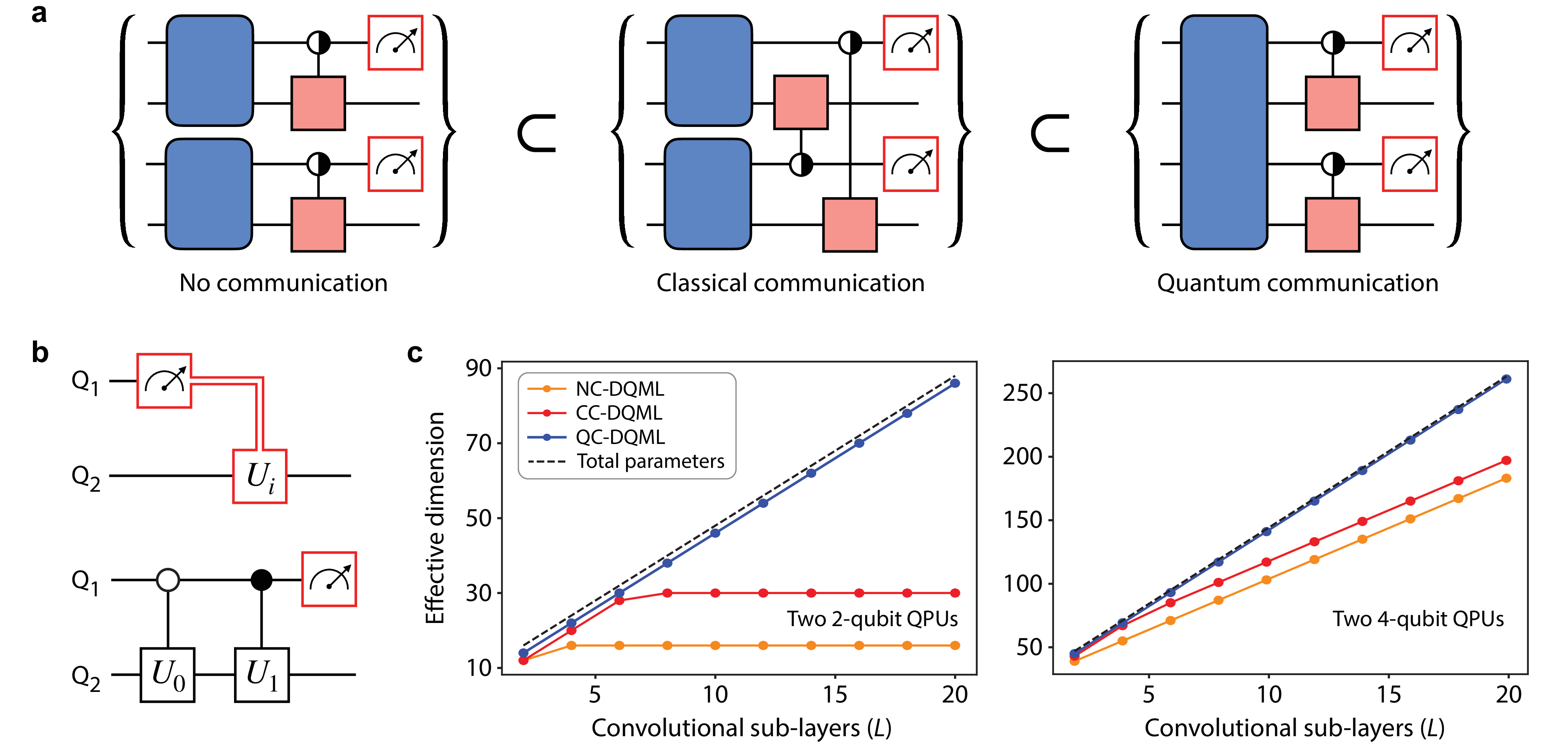}
\caption{\textbf{Circuit capacity analysis.} \textbf{a}, Each set is a collection of implementable circuits corresponding to  no communication, classical communication, and quantum communication schemes. Here, the blue and red blocks represent arbitrary unitary operations and controlled one-qubit operations, respectively, and the symbol $\subset$ indicates their subset relationships. \textbf{b}, Two equivalent circuits: (top) a classical feedforward operation and (bottom) a measurement after two controlled gate operations. \textbf{c}, The effective dimension of using two 2-qubit or 4-qubit QPUs
is estimated for different number of repeated convolutional sub-layers $L$ in Fig.~\ref{fig2}. Note that as $L$ increases, the effective dimension will eventually saturate for all the DQML schemes~\cite{haug2021capacity}. They are parameterized using the same embedding, convolutional, and pooling circuit blocks in Fig.~\ref{fig4}\textbf{b}. 
}
\label{fig3}
\end{figure*}

\begin{figure*}

\centering
\includegraphics[width=\linewidth]{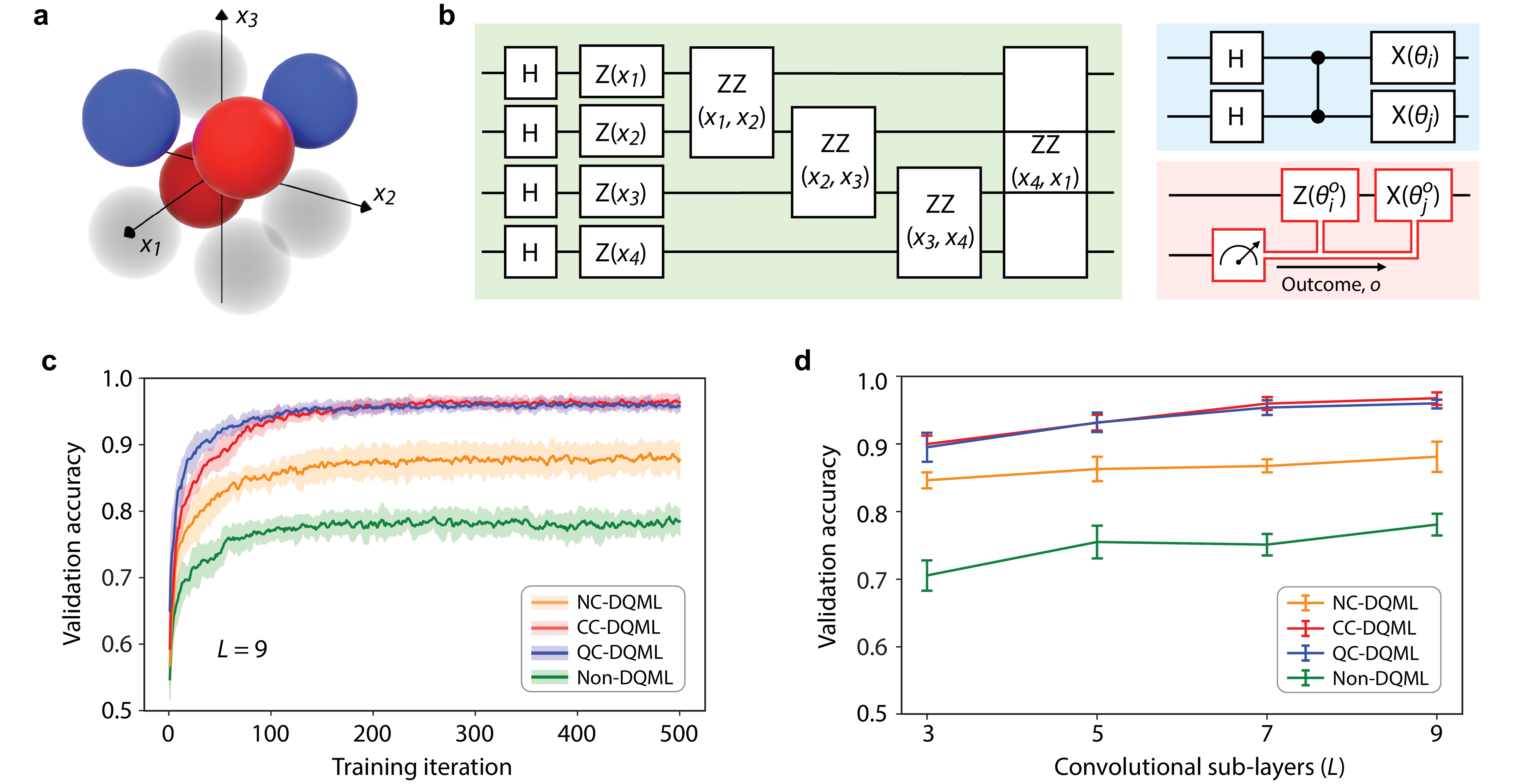}
\caption{\textbf{Binary classification using DQML.} \textbf{a}, A synthetic dataset's attributes, $\boldsymbol{x}=(x_1$, $x_2$, $x_3$), form clusters in a Cartesian space, with blue and red spheres representing different labels for each cluster. Although the plot displays a 3-dimensional dataset, we employ an 8-dimensional dataset for DQML demonstration. 
\textbf{b}, To embed the 8-dimensional dataset, two green blocks are deployed: one block per QPU in DQML, and two blocks for a single QPU in non-DQML. The blue block forms the brick-wall convolutional sub-layers depicted at the bottom of Fig.~\ref{fig2}, while the red block constructs the pooling layers. The number of parameters in each convolutional sub-layer matches the number of qubits involved, with parameter $\theta_i$ assigned based on the corresponding qubit index. Note that the parameters in the red block depend on the measurement outcome $o$, leading to four parameters per block.
\textbf{c, d}, Binary classification of the 8-dimensional dataset is numerically conducted with two 4-qubit QPUs for DQML and a single 4-qubit QPU for non-DQML. The dataset consists of 2048 instances, with 1536 used for training and 512 for validation. Each training iteration processes a batch of 512 instances. See Appendix~\ref{appB} for more details. The results in panels \textbf{c} and \textbf{d} are obtained with $L$=9 convolutional sub-layers and 1000 training iterations, respectively. The solid lines denote the mean, and the shaded regions and error bars represent the standard deviation of 10 trials with random initial parameters. }

\label{fig4}
\end{figure*}

\noindent\textbf{Schematic}\\
Figure~\ref{fig1} illustrates a general schematic of classical communication between quantum processor units (QPUs).
When a mid-circuit measurement is performed on a QPU, it captures partial information about the data in process.
This information is then transferred to another QPU through a conditional feedforward operation, establishing shared classical correlations between them. While quantum machine learning is renowned for exploiting quantum nature, establishing these classical correlations can aid data processing by increasing the utilizable parameter space of a model. 

The classical communication structure mirrors the pooling operation of quantum convolutional neural networks (QCNN)~\cite{cong2019quantum}, involving mid-circuit measurements followed by feedforward operations. Inspired by this similarity, we propose a DQML scheme and the structure is illustrated at the top of Fig.~\ref{fig2}. Training data is loaded into QPUs via embedding layers, and both convolutional layers and pooling layers are parameterized for training. Note that the pooling layers reduce the number of engaged qubits through mid-circuit measurements. 

The DQML schemes can be classified into three categories according to communication methods: no communication (NC), classical communication (CC), and quantum communication (QC). They are depicted at the bottom of Fig.~\ref{fig2}. Classical communication is implemented through feedforward operations
between QPUs, while quantum communication is facilitated by nonlocal two-qubit operations within convolutional sub-layers. In the following sections, we will contrast the CC-DQML scheme with the non-DQML, NC-DQML and QC-DQML schemes to demonstrate the advantages of using classical communication. It is noteworthy that the NC-DQML scheme does not require running on two QPUs simultaneously and can instead be implemented by
combining and post-processing results from two separate runs on a single QPU. \\

\noindent\textbf{Circuit capacity}\\
The performance of a machine learning model hinges on both its circuit capacity and its trainability. More specifically, a model with greater capacity---meaning it can produce more diverse quantum operations---is advantageous as long as it remains trainable. We first compare the circuit capacity of NC-, CC-, and QC-DQML schemes by examining their subset relationships. Given that the two operations in Fig.~\ref{fig3}\textbf{b} are equivalent, we can tell that the sets of implementable circuits in Fig.~\ref{fig3}\textbf{a} correspond to a part of the convolutional and pooling layers of each DQML scheme shown in Fig.~\ref{fig2}. 

In the set of NC circuits, the blue blocks, capable of performing arbitrary unitary operations, make the controlled gates of the red blocks redundant. Moreover, because the red blocks in the set of CC circuits can function as identity operations, we can conclude that the set of NC circuits is a subset of the set of CC circuits. However, this relationship does not apply in reverse because local unitary operations within each processor cannot replace the two-qubit operations between processors. The same logic applies when comparing the sets of CC and QC circuits. This subset relationship can be extended to processors with more than two qubits, demonstrating that classical communication provides greater circuit capacity than no communication, though still less than that offered by quantum communication.

We now quantitatively compare the circuit capacity by analyzing the effective dimension~\cite{abbas2021power}, defined as the maximum rank of the Fisher information matrix over all possible training parameter sets $\boldsymbol{\theta} \in \boldsymbol{\Theta}$:
\begin{equation}
\label{ED}
\text{ED}:=\underset{\boldsymbol{\theta} \in \boldsymbol{\Theta}}{\max}\ r_{\boldsymbol{\theta}},
\end{equation}
where $r_{\boldsymbol{\theta}}$ is the rank of Fisher information matrix for a given parameter setting  $\boldsymbol{\theta}$. The Fisher information quantifies how sensitive an observable or a training loss function is to changes in the training parameters, so the effective dimension reflects the number of independent and significant training parameters. Thereby, the effective dimension offers a quantitative measure of circuit capacity by evaluating the dimensionality of information geometry. Specifically, we obtain the Fisher information matrix defined as:
\begin{equation}
\label{fisher}
F_{ij}(\boldsymbol{\theta})=\frac{1}{N} \sum_{n=1}^{N} \frac{\partial}{\partial \theta_i} \log p(\boldsymbol{x}^{n}, \boldsymbol{y}^{n} ; \boldsymbol{\theta}) \frac{\partial}{\partial \theta_j} \log p(\boldsymbol{x}^{n}, \boldsymbol{y}^{n} ; \boldsymbol{\theta}),
\end{equation}
where $\theta_i$ denotes the $i$-th training parameter of the  convolutional and pooling layers, and $p(\boldsymbol{x}^n,\boldsymbol{y}^n;\boldsymbol{\theta})$ represents the joint probability distribution of the $n$-th data instance $\boldsymbol{x}^n$ and its corresponding output $\boldsymbol{y}^n$.

Figure~\ref{fig3}\textbf{c} presents numerically estimated effective dimensions of DQML schemes with respect to $N=500$ Haar-distributed random data. The maximum rank of the Fisher information is obtained from the ranks calculated using 20 random training parameter sets $\boldsymbol{\theta}$ sampled from uniform distribution. The results suggest that CC-DQML generally exhibits a larger effective dimension and thus greater capacity compared to NC-DQML. Additionally, with a small number of convolutional sub-layers ($L$) that constitute a single convolutional layer in Fig.~\ref{fig2}, we observe that the effective dimension of CC-DQML matches that of QC-DQML, indicating that classical communication can be just as effective as quantum communication at shallow depths. Detailed information on the estimation and further analysis of the effective dimension are available in Appendix~\ref{appA}. 

It is important to clarify that while increasing circuit capacity can enhance the potential for a lower global minimum of training loss functions, the expanded parameter space may hinder trainability, making it more difficult to find the global minimum and achieve fast convergence~\cite{caro2022generalization,haug2021capacity,holmes2022connecting}. However, there is some evidence suggesting that our CC-DQML scheme would exhibit decent trainability. Firstly, the QCNN architecture that inspired our scheme is known for its high trainability and its ability to mitigate exponentially vanishing gradients~\cite{pesah2021absence}. Secondly, the classical approach of splitting circuits is known to result in a larger variance of gradient~\cite{tuysuz2023classical}.\\

\noindent\textbf{Binary classification}\\
We now turn to demonstrate the training performance of DQML schemes through numerical emulation using PennyLane~\cite{bergholm2018pennylane}.
Specifically, we benchmark their classification accuracy on binary classification tasks using 4-qubit QPUs.
Commonly used datasets like MNIST and Iris are easily trainable with even a single 4-qubit QPU~\cite{hur2022quantum, abbas2021power}, so we opt to use 8-dimensional synthetic datasets to better showcase the utility of CC-DQML. The eight attributes of a dataset comprise 32 clusters distributed across an 8-dimensional space, each placed within one of 32 divisions randomly selected from a total of $2^8 = 256$ possible divisions. The binary labels for these clusters are also assigned randomly, intentionally reducing the linear correlation between individual attributes and labels. To provide clarity, a simplified 3-dimensional synthetic dataset is depicted in Fig.~\ref{fig4}\textbf{a}. Further details of the dataset can be found in Appendix~\ref{appB}. The eight attributes are divided into two sets, and each set is assigned to the embedding layer of each QPU. In the case of non-DQML using a single 4-qubit QPU, the embedding layer is repeated twice to accommodate eight attributes onto the four qubits.

For training, we use the DQML models depicted in Fig.~\ref{fig2}, employing the circuit blocks in Fig.~\ref{fig4}\textbf{b}, which have demonstrated high classification accuracy in previous works~\cite{hur2022quantum,sim2019expressibility}. The information of data is compressed by two consecutive pooling layers, and each QPU yields one bit of information through single-qubit measurements (Fig.~\ref{fig2}). To effectively integrate the two-bit information, we adapt interpret function as follows:
\begin{equation}
\label{interpret}
f_\mathrm{int}=w_0 P[00]+w_1 P[01]+w_2 P[10]+w_3 P[11],
\end{equation}
where $P[ij]$ denotes the probability of measuring outcomes
$i$ and $j$ from each QPU, and $\boldsymbol{w}=(w_0,w_1,w_2,w_3)$ represents the associated weights. These weights $w$ are optimized to achieve the best interpretation along with the gate parameters of the convolutional and pooling layers for each training iteration. The least squares loss function for training is defined using the interpret function as $\mathcal{L}\propto\sum_n (f^n_\mathrm{label}-f^n_\mathrm{int})^2$, where $f^n_\mathrm{label}\in\{-1,1\}$ represents the label of the $n$-th data instance. Note that using the interpret function instead of traditional parity measurements significantly improves classification accuracy, especially for NC-DQML and CC-DQML. We attribute this to the lower connectivity of the schemes compared to QC-DQML. See Appendix~\ref{appC} for more details.

Figures~\ref{fig4}\textbf{c} and \textbf{d} exhibit the classification accuracy. With $L$=9 convolutional sub-layers and after 1000 training iterations, both CC-DQML and QC-DQML achieve comparable classification accuracy within a standard deviation, around 96\%, which is significantly higher than the approximately 88\% accuracy of NC-DQML and 78\% of non-DQML. The faster convergence of QC-DQML relative to CC-DQML can be attributed to its lower variance in the non-zero Fisher information spectrum, which reduces gradient vanishing in the optimization landscape~\cite{haug2021capacity,huembeli2021characterizing} (see Appendix~\ref{appD} for the analysis of the Fisher information spectrum). This indicates that quantum communication still holds an advantage over classical communication. Nonetheless, CC-DQML can generate optimal circuits with classification accuracy similar to those of QC-DQML with various number of sub-layers, and this implies that, in some cases like ours, classical communication can be just as effective as quantum communication. To demonstrate that these results are not limited to a single dataset, we have included additional training results with other datasets in Appendix~\ref{appB}, along with the results using 15 and 20 convolutional sub-layers.\\



\noindent\textbf{Discussion}\\
We have introduced a practical and effective DQML scheme for processing larger-scale data by integrating separate quantum processor units. Our numerical results in Fig.~\ref{fig4}\textbf{d} demonstrate that DQML schemes with $L$=3 convolutional sub-layers achieve higher classification accuracy than a non-DQML scheme with $L$=7, despite similar number of training parameters. Especially, NC-DQML's superior performance highlights the benefits of distributed approaches, whether across multiple processors or multiple runs within a single processor.

The similar performance of CC-DQML and QC-DQML suggests that classical communication can serve as a viable alternative to quantum communication. While QC-DQML may outperform with datasets featuring highly correlated attributes or requiring extensive parameter dimensionality, classical communication offers a feasible solution given current technology constraints in the noisy intermediate-scale Quantum (NISQ) era. Further research is needed to optimize CC-DQML strategies for data division and embedding, as well as to identify the best interpret function for post-processing outcomes from each processor. In addition, exploring DQML schemes with more than two processors could reveal further advantages and applications. We hope our work provides a practical pathway for handling high-dimensional datasets and expanding the usable quantum space, helping to overcome current limitations in qubit availability.\\

\noindent\textbf{Acknowledgements}\\
This work was supported by the National Research Foundation of Korea (NRF) grant funded by the Korea government (MSIT) (Nos. RS-2024-00432563, RS-2024-00404854, 2022M3E4A1043330, 2022M3K4A1097123, 2023M3K5A1094805, and 2023M3K5A1094813), Institute for Information \& communications Technology Promotion (IITP) grant funded by the Korea government (MSIT) (No. 2019-0-00003), the Yonsei University Research Fund (No. 2024-22-0147), and the KIST research programs (Nos. 2E32941 and 2E32971).

\newpage

\appendix

\section{Effective dimension estimation}\label{appA}
To assess the general characteristics of our DQML models in Fig.~\ref{fig2}, we estimate their effective dimension using Haar-distributed random datasets. For this estimation, the embedding layer of each QPU is constructed using Haar random unitaries, while the remaining circuit blocks are configured as illustrated in Fig.~\ref{fig4}\textbf{b}. Note that, to obtain the results presented in Fig.~\ref{fig3}\textbf{c}, a single pooling layer is applied for two 2-qubit QPUs, whereas two pooling layers are used for two 4-qubit QPUs. The first step in estimating the effective dimension involves calculating the Fisher information matrices of Eq.~(\ref{fisher}) for a dataset consisting of 500 Haar random quantum states. These calculations are performed across 20 different sets of training parameters, sampled from a uniform distribution, $\boldsymbol{\theta} \in [0,2\pi)^d$, where $d$ represents the number of total parameters in a model. Subsequently, the rank of each Fisher information matrix is evaluated for each set of training parameters, and the effective dimension is determined by taking the maximum rank obtained. Although, as defined by Eq.~(\ref{ED}), the effective dimension should represent the highest rank attainable across all possible training parameter settings, our findings suggest that 20 random parameter sets are sufficient to produce consistent results. 

\begin{figure}[t!]
\centering
\includegraphics[width=\linewidth]{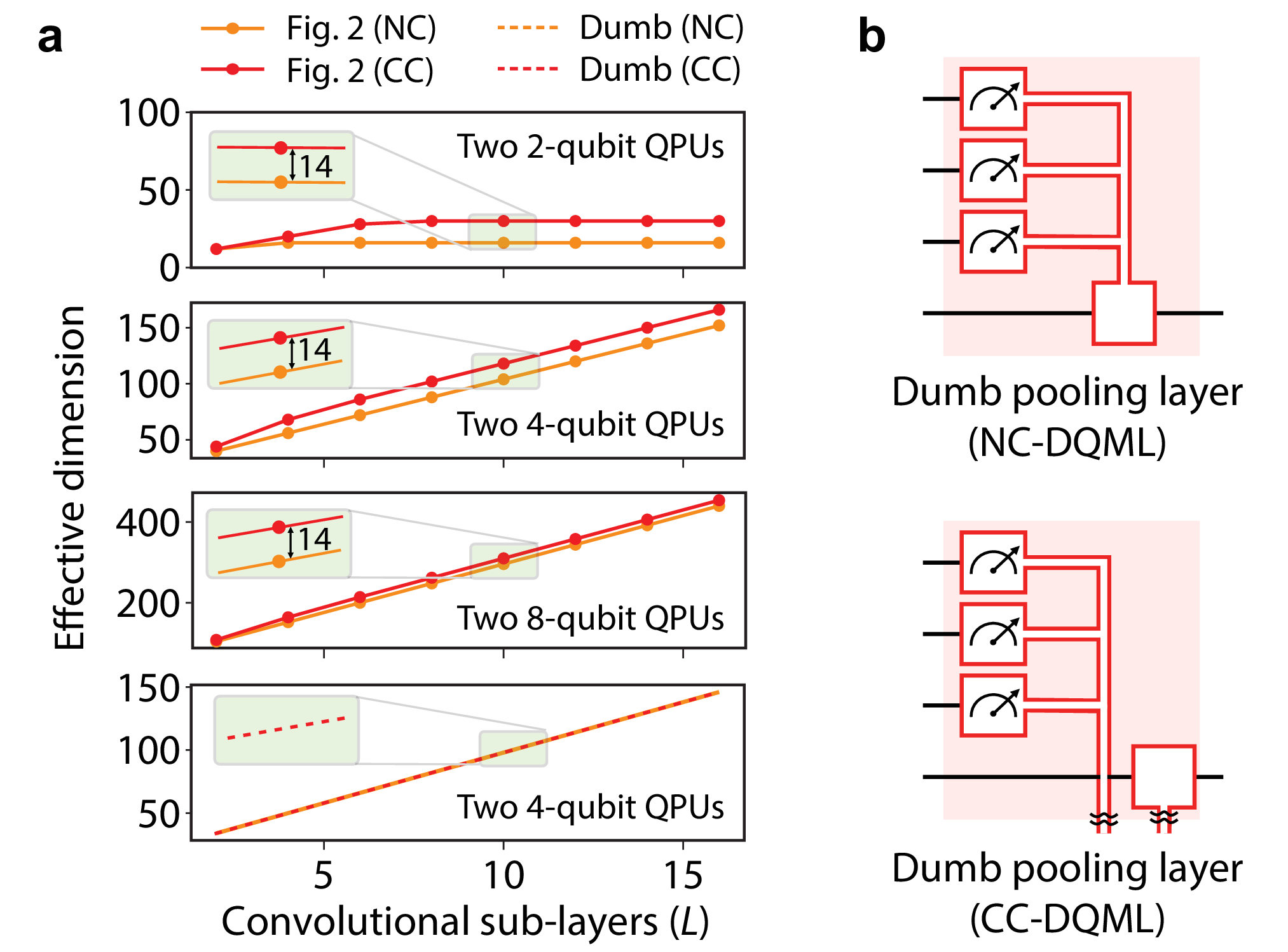}
\caption{\textbf{Comparison of NC-DQML and CC-DQML.} \textbf{a}, The top three plots display the effective dimension of NC-DQML and CC-DQML using two QPUs with varying qubit counts. These estimations are based on the DQML models shown in Figs.~\ref{fig2}~and~\ref{fig4}\textbf{b}, utilizing Haar random datasets. The bottom plot presents the effective dimension of the simplified pooling layer. \textbf{b}, The simplified dumb pooling layer is introduced to explore how the structure of the pooling layer affects the effective dimension. In each QPU, all qubits except one are measured and a feedforward operation applied to the remaining qubit, similar to the red block shown in Fig.~\ref{fig4}\textbf{b}. 
Specifically, three pairs of Z and X gates are applied to each qubit. The pooling layer reduces the number of qubits to one, so the DQML models consist of a single convolutional layer followed by a pooling layer when using two 4-qubit QPUs.
}
\label{fig5}
\end{figure}

In our models, we observe an interesting trend in the gap between the effective dimensions of NC-DQML and CC-DQML. As can be seen in Fig.~\ref{fig5}\textbf{a}, this gap consistently remains at 14 across different numbers of convolutional sub-layers ($L$) even as the number of qubits in the QPUs increases from 2 to 8. However, we want to emphasize that this does not imply  classical communication offers only marginal benefits. In fact, CC-DQML's effective dimension reaches saturation at a higher $L$, leading to a larger difference at $L$ where NC-DQML's effective dimension has already saturated. For example, with two 4-qubit QPUs and $L=30$ convolutional sub-layers, we find that CC-DQML achieves an effective dimension of 278, whereas NC-DQML reaches 256. Additionally, this gap can vary depending on the pooling structure. For instance, the plot at the bottom of Fig.~\ref{fig5}\textbf{a}, which shows results for NC-DQML and QC-DQML with pooling layers in Fig.~\ref{fig5}\textbf{b}, indicates no difference between them. This suggests that further research is needed to fully understand and optimize the benefits of classical communication, especially in relation to the effective dimension gap.

\begin{table}[t!]
\centering
\begin{tabular}{      c||c|c|c|c  }

 \hline

 \multicolumn{5}{c}{Validation accuracy (\%)} \\
 \hline
 \multirow{2}{*}{$L$}  &   \multirow{2}{*}{non-DQML}   & \multirow{2}{*}{NC-DQML}  & \multirow{2}{*}{CC-DQML} & \multirow{2}{*}{QC-DQML}\\
 &&&&\\

 \hline

 \multirow{3}{*}{3} &  70.57 $\pm$ 2.31  & 84.63 $\pm$ 1.28   & 89.98 $\pm$ 1.19 & 89.51 $\pm$ 2.07 \\[1pt]
 & 71.23 $\pm$ 1.38 & 85.68 $\pm$ 2.35 & 86.74 $\pm$ 1.83 & 88.30 $\pm$ 1.67\\[1pt]
 &  71.43 $\pm$ 2.24  & 86.97 $\pm$ 1.91   & 89.36 $\pm$ 1.07 & 87.13 $\pm$ 1.89\\

 \hline
 \multirow{3}{*}{5}  &75.51 $\pm$ 2.48 & 86.29 $\pm$ 1.92 & 93.12 $\pm$ 1.26 & 93.16 $\pm$ 1.50\\[1pt]
 &74.04 $\pm$ 1.69 &88.59 $\pm$ 3.04 &91.05 $\pm$ 1.80 &91.89 $\pm$ 1.29\\[1pt]

 &75.27 $\pm$ 1.72 & 90.92 $\pm$ 1.31 & 91.88 $\pm$ 1.80 & 91.68 $\pm$ 0.82\\

\hline

 \multirow{3}{*}{7}
 & 75.12 $\pm$ 1.66 & 86.74 $\pm$ 1.05 & 95.96 $\pm$ 1.06 & 95.37 $\pm$ 1.17\\[1pt]
 &75.18 $\pm$ 2.29 & 88.79 $\pm$ 2.22 & 95.25 $\pm$ 1.28 & 94.22 $\pm$ 0.93\\[1pt]
& 76.88 $\pm$ 2.05 & 90.96 $\pm$ 0.94 & 94.43 $\pm$ 0.79 & 93.67 $\pm$ 1.11\\

 \hline
 \multirow{3}{*}{9}&78.09 $\pm$ 1.68 &88.11 $\pm$ 2.33 &96.76 $\pm$ 0.97 &96.00 $\pm$ 0.81\\[1pt]
 & 77.27 $\pm$ 1.33 & 89.49 $\pm$ 2.76 & 96.00 $\pm$ 1.31 & 95.94 $\pm$ 0.85\\[1pt]
&78.24 $\pm$ 1.31 &92.25 $\pm$ 1.53&96.52$\pm$ 0.98&95.70 $\pm$ 0.86\\
\hline

15 &81.66 $\pm$ 1.38 &87.58 $\pm$ 1.98 &98.63 $\pm$ 0.51 &98.01 $\pm$ 0.61\\[1pt]

\hline

 20 &83.63 $\pm$ 0.73 &88.22 $\pm$ 1.35 &99.41 $\pm$ 0.36 &99.08 $\pm$ 0.41\\[1pt]

\hline
\end{tabular}
\caption{\textbf{Binary classification accuracy.} The results are obtained using the (non-)DQML models illustrated in Figs.~\ref{fig2}~and~\ref{fig4}\textbf{b}.  For the non-DQML scheme, a single 4-qubit QPU is used, while two QPUs are employed for the three DQML schemes. The mean and standard deviation are calculated at 1000 training iterations from 10 trials with random initial parameters. The results listed in the first row for each $L$ correspond to those shown in Figs.~\ref{fig4}\textbf{c} and \textbf{d}, and the results for $L=15$ and $L=20$ are obtained from the same dataset.
}
\label{tab1}
\end{table}

\begin{table}[h!]
\centering
\begin{tabular}{      c||c|c|c  }

 \hline

 \multicolumn{4}{c}{Validation accuracy (\%)} \\
 \hline
 \multirow{2}{*}{$L$}  & \multirow{2}{*}{NC-DQML}  & \multirow{2}{*}{CC-DQML} & \multirow{2}{*}{QC-DQML}\\
 &&&\\

 \hline

 \multirow{2}{*}{3} & 75.55 $\pm$ 2.87 & 82.54 $\pm$ 1.94 & 81.52 $\pm$ 3.26 \\[1pt]
  &($\Delta=9.08$)&($\Delta=7.44$)&($\Delta=7.99$)\\

 \hline
 \multirow{2}{*}{5}   & 79.84 $\pm$ 1.90 & 82.7 $\pm$ 2.71 & 88.75 $\pm$ 1.08 \\[1pt]
 &($\Delta=6.45$)&($\Delta=10.43$)&($\Delta=4.41$)\\
 
\hline

 \multirow{2}{*}{7}
& 83.52 $\pm$ 2.97 & 87.07 $\pm$ 3.03 & 91.21 $\pm$ 1.72 \\[1pt]
&($\Delta=3.22$)&($\Delta=8.89$)&($\Delta=4.16$)\\

 \hline
 \multirow{2}{*}{9} & 85.00 $\pm$ 2.58 & 87.23 $\pm$ 2.41 & 92.81 $\pm$ 0.42 \\[1pt]
&($\Delta=3.11$)&($\Delta=9.53$)&($\Delta=3.18$)\\

\hline
\end{tabular}
\caption{\textbf{Effectiveness of the interpret function.} The results are obtained using the DQML models illustrated in Figs.~\ref{fig2}~and~\ref{fig4}\textbf{b}. Unlike the results in Table\ref{tab1}, which employ the interpret function, here the measurement outcomes are post-processed using the parity interpretation. This significantly reduces classification accuracy. $\Delta$ denotes the difference compared to the results obtained with the interpret function. The mean and standard deviation are calculated at 1000 training iterations from 5 trials with random initial parameters, using the dataset employed for Figs.~\ref{fig4}\textbf{c} and \textbf{d}.}
\label{tab2}
\end{table}

\section{Dataset synthesis and classification}\label{appB}

To demonstrate the effectiveness of DQML, we generate synthetic datasets having a low linear correlation between each attribute and binary labels, which are designed to be challenging for binary classification tasks. For an 8-dimensional dataset, we start by uniformly sampling 2048 vectors, $\boldsymbol{x}=(x_1, x_2, ..., x_8)$, from within an 8-dimensional sphere of radius $\pi/4$: $\{\boldsymbol{x}|\sum_{i=1}^8 x_i^2<(\pi/4)^2\}$. Next, we randomly form 32 clusters and assign binary labels $f_\mathrm{label}\in\{-1,1\}$ to each cluster, ensuring that there is an equal number of -1 and 1 labels across all clusters. Each cluster contains 64 data instances, since 2048/32=64, and it is translated by randomly selected vectors $\boldsymbol{v}=(v_1,v_2,...,v_7,v_8)$ where $v_i\in\{-\pi/4,\pi/4\}$. Among the 2048 instances, 1536 instances are allocated to the training set and the remaining 512 instances are used for the validation set. To evaluate the linear correlation between individual attribute and the label, we calculate the Pearson correlation coefficient which is defined as $\rho_i=\mathrm{cov}(x_i,f_\mathrm{label})/\sigma_{x_i}\sigma_{f}$. Here, $\mathrm{cov}(x_i,f_\mathrm{label})$ represents the covariance of the attribute $x_i$ and the label $f_\mathrm{label}$, and $\sigma_{x_i}$ and $\sigma_{f}$ denote the standard deviations of the attribute and the labels, respectively, for the 2048 instances. The maximum correlation is estimated to be $\rho_\mathrm{max}=\mathrm{max}( \rho_i)=0.239$ for the dataset used for Figs.~\ref{fig4}\textbf{c} and \textbf{d}. This indicates that its binary classification would be challenging with only a subset of the attributes.

We use synthetic datasets to numerically demonstrate binary classification using the PennyLane library~\cite{bergholm2018pennylane}. The Adam optimizer is configured with a step size of 0.05~\cite{kingma2014adam}, and each training iteration processes a batch of 512 instances. Initial training parameters for the convolutional and pooling layers are drawn from a uniform distribution $\boldsymbol{\theta} \in [0,2\pi)^d$, while the initial parameters for the interpret function in Eq.~(\ref{interpret}) is fixed to $\boldsymbol{w}=(1,-1,-1,1)$, which corresponds to the parity measurements.
Table~\ref{tab1} presents the classification accuracy for the validation set using 4-qubit QPUs. Three datasets are tested, and each dataset was randomly generated using the previously described method. The results indicate that CC-DQML consistently outperforms both NC-DQML and non-DQML, and is competitive with QC-DQML for all the datasets.

\begin{figure}[t!]
\centering
\includegraphics[width=\linewidth]{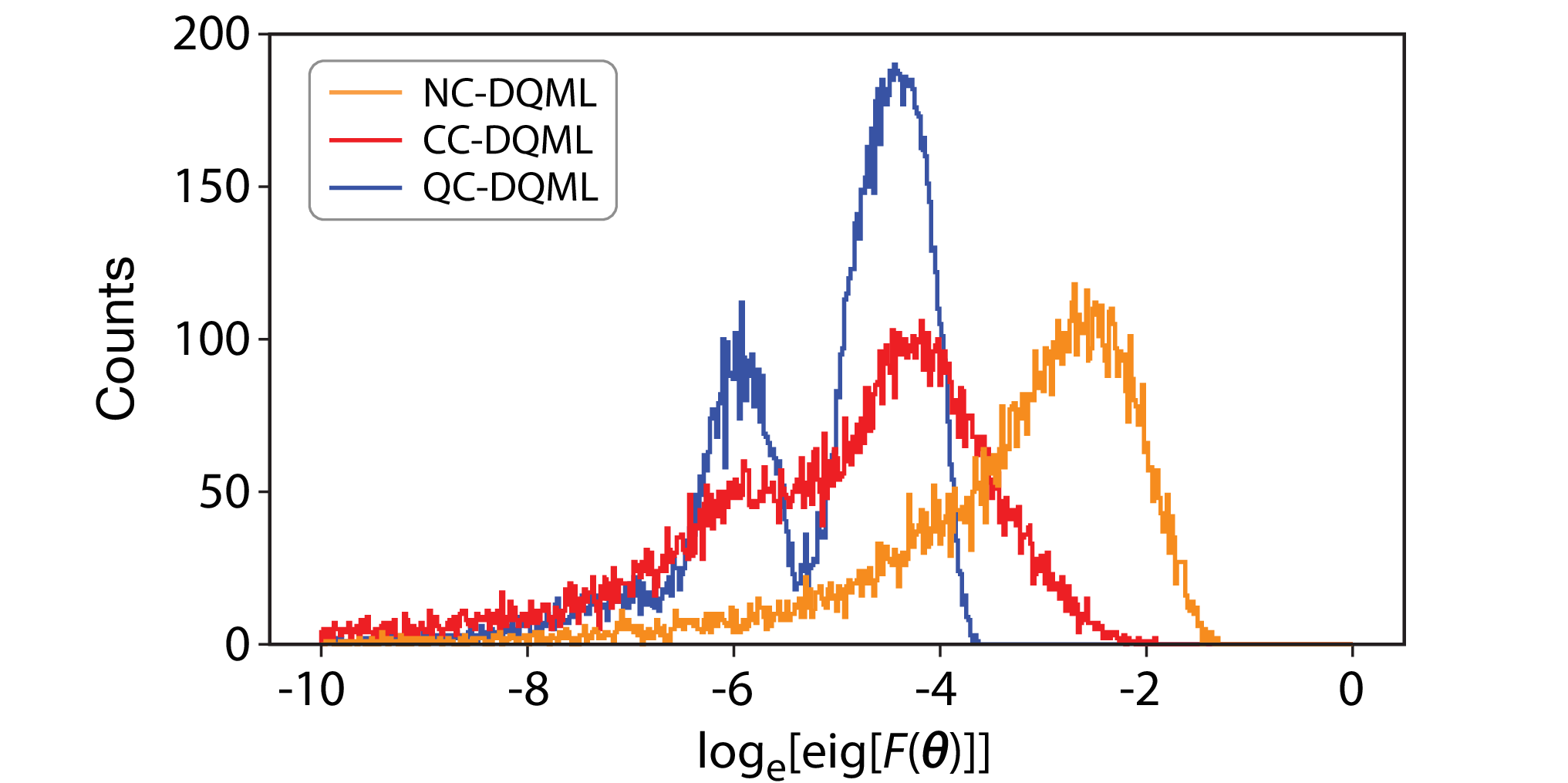}
\caption{\textbf{Histogram of the Fisher information spectrum.} The histogram depicts the non-zero eigenvalue spectra of Fisher information matrices on a log scale for DQML schemes utilizing two 4-qubit QPUs with $L=4$ convolutional sub-layers. To avoid any potential bias, the histogram is obtained by accumulating the eigenvalues of 200 Fisher information matrices with 200 random training parameter sets, $\boldsymbol{\theta}$. Each Fisher information matrix is estimated using 200 Haar random quantum states as a dataset. The logarithmic bin width of the histogram is set to 0.02, and the standard deviations of the logarithms of the non-zero eigenvalues for NC-DQML, CC-DQML, and QC-DQML are estimated to be 1.383, 2.136, and 1.056, respectively.
}
\label{fig6}
\end{figure}

\section{Interpret function}\label{appC}
Quantum communication allows for universal quantum operations between separate processors, enabling the generation of any desired correlations for data processing. In contrast, classical communication can only produce limited correlations, which may reduce its effectiveness. To overcome this limitation, we introduce the interpret function to post-process the measurement outcomes, expressed as $w_0 P[00]+w_1 P[01]+w_2 P[10]+w_3 P[11]$, where $P[ij]$ denotes the probability of measuring outcomes $i$ and $j$ from each QPU, and $\boldsymbol{w}=(w_0,w_1,w_2,w_3)$ represents the associated weights. These weights are optimized alongside the training parameters of the convolutional and pooling layers to obtain the best possible interpretation. Table~\ref{tab2} displays the classification accuracy of DQML schemes using the parity interpretation, ($P[00]- P[01]- P[10]+P[11]$), and the difference in accuracy with the results obtained using the interpret function. With the parity interpretation, CC-DQML achieves higher classification accuracy than NC-DQML but does not surpass QC-DQML. Given that CC-DQML shows comparable validation accuracy to QC-DQML in Table~\ref{tab1}, when using the interpret function, it highlights the effectiveness of the interpret function.

\section{Fisher information spectrum}\label{appD}

Fisher information spectrum offers a glimpse into the overall optimization landscape of a training model. While the effective dimension assesses capacity of a quantum circuit by examining the degeneracy of the Fisher information spectrum, the distribution of non-zero eigenvalues foreshadows quality of guidance a model offers.
If the non-zero eigenvalues of a Fisher information matrix are concentrated near zero, the model is more likely to have poor trainability, potentially including the presence of a barren plateau~\cite{haug2021capacity,huembeli2021characterizing}. Note that small eigenvalues indicate small gradients in general during the learning process. To compare the Fisher information spectra of DQML schemes, we obtain the histogram of non-zero eigenvalues using Haar random quantum states (Fig.~\ref{fig6}). Since the relative values of non-zero eigenvalues are crucial for understanding the overall landscape, we plot the histogram on a log scale. The standard deviations of the logarithms of the non-zero eigenvalues for NC-, CC-, and QC-DQML are estimated to be 1.383, 2.136, and 1.056, respectively. This suggests that CC-DQML has  poorer trainability than QC-DQML, which explains the slower convergence of CC-DQML compared to QC-DQML, as can be seen in Fig.~\ref{fig4}\textbf{c}.

\bigskip

\bibliography{Bib}

\begin{thebibliography}{48}%
\makeatletter
\providecommand \@ifxundefined [1]{%
 \@ifx{#1\undefined}
}%
\providecommand \@ifnum [1]{%
 \ifnum #1\expandafter \@firstoftwo
 \else \expandafter \@secondoftwo
 \fi
}%
\providecommand \@ifx [1]{%
 \ifx #1\expandafter \@firstoftwo
 \else \expandafter \@secondoftwo
 \fi
}%
\providecommand \natexlab [1]{#1}%
\providecommand \enquote  [1]{``#1''}%
\providecommand \bibnamefont  [1]{#1}%
\providecommand \bibfnamefont [1]{#1}%
\providecommand \citenamefont [1]{#1}%
\providecommand \href@noop [0]{\@secondoftwo}%
\providecommand \href [0]{\begingroup \@sanitize@url \@href}%
\providecommand \@href[1]{\@@startlink{#1}\@@href}%
\providecommand \@@href[1]{\endgroup#1\@@endlink}%
\providecommand \@sanitize@url [0]{\catcode `\\12\catcode `\$12\catcode
  `\&12\catcode `\#12\catcode `\^12\catcode `\_12\catcode `\%12\relax}%
\providecommand \@@startlink[1]{}%
\providecommand \@@endlink[0]{}%
\providecommand \url  [0]{\begingroup\@sanitize@url \@url }%
\providecommand \@url [1]{\endgroup\@href {#1}{\urlprefix }}%
\providecommand \urlprefix  [0]{URL }%
\providecommand \Eprint [0]{\href }%
\providecommand \doibase [0]{https://doi.org/}%
\providecommand \selectlanguage [0]{\@gobble}%
\providecommand \bibinfo  [0]{\@secondoftwo}%
\providecommand \bibfield  [0]{\@secondoftwo}%
\providecommand \translation [1]{[#1]}%
\providecommand \BibitemOpen [0]{}%
\providecommand \bibitemStop [0]{}%
\providecommand \bibitemNoStop [0]{.\EOS\space}%
\providecommand \EOS [0]{\spacefactor3000\relax}%
\providecommand \BibitemShut  [1]{\csname bibitem#1\endcsname}%
\let\auto@bib@innerbib\@empty
\bibitem [{\citenamefont {Shor}(1999)}]{shor1999polynomial}%
  \BibitemOpen
  \bibfield  {author} {\bibinfo {author} {\bibfnamefont {P.~W.}\ \bibnamefont
  {Shor}},\ }\bibfield  {title} {\bibinfo {title} {Polynomial-time algorithms
  for prime factorization and discrete logarithms on a quantum computer},\
  }\href@noop {} {\bibfield  {journal} {\bibinfo  {journal} {SIAM review}\
  }\textbf {\bibinfo {volume} {41}},\ \bibinfo {pages} {303} (\bibinfo {year}
  {1999})}\BibitemShut {NoStop}%
\bibitem [{\citenamefont {Grover}(1996)}]{grover1996fast}%
  \BibitemOpen
  \bibfield  {author} {\bibinfo {author} {\bibfnamefont {L.~K.}\ \bibnamefont
  {Grover}},\ }\bibfield  {title} {\bibinfo {title} {A fast quantum mechanical
  algorithm for database search},\ }in\ \href@noop {} {\emph {\bibinfo
  {booktitle} {Proceedings of the twenty-eighth annual ACM symposium on Theory
  of computing}}}\ (\bibinfo {year} {1996})\ pp.\ \bibinfo {pages}
  {212--219}\BibitemShut {NoStop}%
\bibitem [{\citenamefont {Arute}\ \emph {et~al.}(2019)\citenamefont {Arute},
  \citenamefont {Arya}, \citenamefont {Babbush}, \citenamefont {Bacon},
  \citenamefont {Bardin}, \citenamefont {Barends}, \citenamefont {Biswas},
  \citenamefont {Boixo}, \citenamefont {Brandao}, \citenamefont {Buell} \emph
  {et~al.}}]{arute2019quantum}%
  \BibitemOpen
  \bibfield  {author} {\bibinfo {author} {\bibfnamefont {F.}~\bibnamefont
  {Arute}}, \bibinfo {author} {\bibfnamefont {K.}~\bibnamefont {Arya}},
  \bibinfo {author} {\bibfnamefont {R.}~\bibnamefont {Babbush}}, \bibinfo
  {author} {\bibfnamefont {D.}~\bibnamefont {Bacon}}, \bibinfo {author}
  {\bibfnamefont {J.~C.}\ \bibnamefont {Bardin}}, \bibinfo {author}
  {\bibfnamefont {R.}~\bibnamefont {Barends}}, \bibinfo {author} {\bibfnamefont
  {R.}~\bibnamefont {Biswas}}, \bibinfo {author} {\bibfnamefont
  {S.}~\bibnamefont {Boixo}}, \bibinfo {author} {\bibfnamefont {F.~G.}\
  \bibnamefont {Brandao}}, \bibinfo {author} {\bibfnamefont {D.~A.}\
  \bibnamefont {Buell}}, \emph {et~al.},\ }\bibfield  {title} {\bibinfo {title}
  {Quantum supremacy using a programmable superconducting processor},\
  }\href@noop {} {\bibfield  {journal} {\bibinfo  {journal} {Nature}\ }\textbf
  {\bibinfo {volume} {574}},\ \bibinfo {pages} {505} (\bibinfo {year}
  {2019})}\BibitemShut {NoStop}%
\bibitem [{\citenamefont {Pan}\ \emph {et~al.}(2022)\citenamefont {Pan},
  \citenamefont {Chen},\ and\ \citenamefont {Zhang}}]{pan2022solving}%
  \BibitemOpen
  \bibfield  {author} {\bibinfo {author} {\bibfnamefont {F.}~\bibnamefont
  {Pan}}, \bibinfo {author} {\bibfnamefont {K.}~\bibnamefont {Chen}},\ and\
  \bibinfo {author} {\bibfnamefont {P.}~\bibnamefont {Zhang}},\ }\bibfield
  {title} {\bibinfo {title} {Solving the sampling problem of the sycamore
  quantum circuits},\ }\href@noop {} {\bibfield  {journal} {\bibinfo  {journal}
  {Physical Review Letters}\ }\textbf {\bibinfo {volume} {129}},\ \bibinfo
  {pages} {090502} (\bibinfo {year} {2022})}\BibitemShut {NoStop}%
\bibitem [{\citenamefont {Kim}\ \emph {et~al.}(2023)\citenamefont {Kim},
  \citenamefont {Eddins}, \citenamefont {Anand}, \citenamefont {Wei},
  \citenamefont {Van Den~Berg}, \citenamefont {Rosenblatt}, \citenamefont
  {Nayfeh}, \citenamefont {Wu}, \citenamefont {Zaletel}, \citenamefont {Temme}
  \emph {et~al.}}]{kim2023evidence}%
  \BibitemOpen
  \bibfield  {author} {\bibinfo {author} {\bibfnamefont {Y.}~\bibnamefont
  {Kim}}, \bibinfo {author} {\bibfnamefont {A.}~\bibnamefont {Eddins}},
  \bibinfo {author} {\bibfnamefont {S.}~\bibnamefont {Anand}}, \bibinfo
  {author} {\bibfnamefont {K.~X.}\ \bibnamefont {Wei}}, \bibinfo {author}
  {\bibfnamefont {E.}~\bibnamefont {Van Den~Berg}}, \bibinfo {author}
  {\bibfnamefont {S.}~\bibnamefont {Rosenblatt}}, \bibinfo {author}
  {\bibfnamefont {H.}~\bibnamefont {Nayfeh}}, \bibinfo {author} {\bibfnamefont
  {Y.}~\bibnamefont {Wu}}, \bibinfo {author} {\bibfnamefont {M.}~\bibnamefont
  {Zaletel}}, \bibinfo {author} {\bibfnamefont {K.}~\bibnamefont {Temme}},
  \emph {et~al.},\ }\bibfield  {title} {\bibinfo {title} {Evidence for the
  utility of quantum computing before fault tolerance},\ }\href@noop {}
  {\bibfield  {journal} {\bibinfo  {journal} {Nature}\ }\textbf {\bibinfo
  {volume} {618}},\ \bibinfo {pages} {500} (\bibinfo {year}
  {2023})}\BibitemShut {NoStop}%
\bibitem [{\citenamefont {Tindall}\ \emph {et~al.}(2024)\citenamefont
  {Tindall}, \citenamefont {Fishman}, \citenamefont {Stoudenmire},\ and\
  \citenamefont {Sels}}]{tindall2024efficient}%
  \BibitemOpen
  \bibfield  {author} {\bibinfo {author} {\bibfnamefont {J.}~\bibnamefont
  {Tindall}}, \bibinfo {author} {\bibfnamefont {M.}~\bibnamefont {Fishman}},
  \bibinfo {author} {\bibfnamefont {M.}~\bibnamefont {Stoudenmire}},\ and\
  \bibinfo {author} {\bibfnamefont {D.}~\bibnamefont {Sels}},\ }\bibfield
  {title} {\bibinfo {title} {Efficient tensor network simulation of {IBM}'s
  kicked ising experiment},\ }\href@noop {} {\bibfield  {journal} {\bibinfo
  {journal} {PRX Quantum}\ }\textbf {\bibinfo {volume} {5}},\ \bibinfo {pages}
  {010308} (\bibinfo {year} {2024})}\BibitemShut {NoStop}%
\bibitem [{\citenamefont {Kimble}(2008)}]{kimble2008quantum}%
  \BibitemOpen
  \bibfield  {author} {\bibinfo {author} {\bibfnamefont {H.~J.}\ \bibnamefont
  {Kimble}},\ }\bibfield  {title} {\bibinfo {title} {The quantum internet},\
  }\href@noop {} {\bibfield  {journal} {\bibinfo  {journal} {Nature}\ }\textbf
  {\bibinfo {volume} {453}},\ \bibinfo {pages} {1023} (\bibinfo {year}
  {2008})}\BibitemShut {NoStop}%
\bibitem [{\citenamefont {Monroe}\ \emph {et~al.}(2014)\citenamefont {Monroe},
  \citenamefont {Raussendorf}, \citenamefont {Ruthven}, \citenamefont {Brown},
  \citenamefont {Maunz}, \citenamefont {Duan},\ and\ \citenamefont
  {Kim}}]{monroe2014large}%
  \BibitemOpen
  \bibfield  {author} {\bibinfo {author} {\bibfnamefont {C.}~\bibnamefont
  {Monroe}}, \bibinfo {author} {\bibfnamefont {R.}~\bibnamefont {Raussendorf}},
  \bibinfo {author} {\bibfnamefont {A.}~\bibnamefont {Ruthven}}, \bibinfo
  {author} {\bibfnamefont {K.~R.}\ \bibnamefont {Brown}}, \bibinfo {author}
  {\bibfnamefont {P.}~\bibnamefont {Maunz}}, \bibinfo {author} {\bibfnamefont
  {L.-M.}\ \bibnamefont {Duan}},\ and\ \bibinfo {author} {\bibfnamefont
  {J.}~\bibnamefont {Kim}},\ }\bibfield  {title} {\bibinfo {title} {Large-scale
  modular quantum-computer architecture with atomic memory and photonic
  interconnects},\ }\href@noop {} {\bibfield  {journal} {\bibinfo  {journal}
  {Physical Review A}\ }\textbf {\bibinfo {volume} {89}},\ \bibinfo {pages}
  {022317} (\bibinfo {year} {2014})}\BibitemShut {NoStop}%
\bibitem [{\citenamefont {Bravyi}\ \emph {et~al.}(2022)\citenamefont {Bravyi},
  \citenamefont {Dial}, \citenamefont {Gambetta}, \citenamefont {Gil},\ and\
  \citenamefont {Nazario}}]{bravyi2022future}%
  \BibitemOpen
  \bibfield  {author} {\bibinfo {author} {\bibfnamefont {S.}~\bibnamefont
  {Bravyi}}, \bibinfo {author} {\bibfnamefont {O.}~\bibnamefont {Dial}},
  \bibinfo {author} {\bibfnamefont {J.~M.}\ \bibnamefont {Gambetta}}, \bibinfo
  {author} {\bibfnamefont {D.}~\bibnamefont {Gil}},\ and\ \bibinfo {author}
  {\bibfnamefont {Z.}~\bibnamefont {Nazario}},\ }\bibfield  {title} {\bibinfo
  {title} {The future of quantum computing with superconducting qubits},\
  }\href@noop {} {\bibfield  {journal} {\bibinfo  {journal} {Journal of Applied
  Physics}\ }\textbf {\bibinfo {volume} {132}} (\bibinfo {year}
  {2022})}\BibitemShut {NoStop}%
\bibitem [{\citenamefont {Peng}\ \emph {et~al.}(2020)\citenamefont {Peng},
  \citenamefont {Harrow}, \citenamefont {Ozols},\ and\ \citenamefont
  {Wu}}]{peng2020simulating}%
  \BibitemOpen
  \bibfield  {author} {\bibinfo {author} {\bibfnamefont {T.}~\bibnamefont
  {Peng}}, \bibinfo {author} {\bibfnamefont {A.~W.}\ \bibnamefont {Harrow}},
  \bibinfo {author} {\bibfnamefont {M.}~\bibnamefont {Ozols}},\ and\ \bibinfo
  {author} {\bibfnamefont {X.}~\bibnamefont {Wu}},\ }\bibfield  {title}
  {\bibinfo {title} {Simulating large quantum circuits on a small quantum
  computer},\ }\href@noop {} {\bibfield  {journal} {\bibinfo  {journal}
  {Physical review letters}\ }\textbf {\bibinfo {volume} {125}},\ \bibinfo
  {pages} {150504} (\bibinfo {year} {2020})}\BibitemShut {NoStop}%
\bibitem [{\citenamefont {Eddins}\ \emph {et~al.}(2022)\citenamefont {Eddins},
  \citenamefont {Motta}, \citenamefont {Gujarati}, \citenamefont {Bravyi},
  \citenamefont {Mezzacapo}, \citenamefont {Hadfield},\ and\ \citenamefont
  {Sheldon}}]{eddins2022doubling}%
  \BibitemOpen
  \bibfield  {author} {\bibinfo {author} {\bibfnamefont {A.}~\bibnamefont
  {Eddins}}, \bibinfo {author} {\bibfnamefont {M.}~\bibnamefont {Motta}},
  \bibinfo {author} {\bibfnamefont {T.~P.}\ \bibnamefont {Gujarati}}, \bibinfo
  {author} {\bibfnamefont {S.}~\bibnamefont {Bravyi}}, \bibinfo {author}
  {\bibfnamefont {A.}~\bibnamefont {Mezzacapo}}, \bibinfo {author}
  {\bibfnamefont {C.}~\bibnamefont {Hadfield}},\ and\ \bibinfo {author}
  {\bibfnamefont {S.}~\bibnamefont {Sheldon}},\ }\bibfield  {title} {\bibinfo
  {title} {Doubling the size of quantum simulators by entanglement forging},\
  }\href@noop {} {\bibfield  {journal} {\bibinfo  {journal} {PRX Quantum}\
  }\textbf {\bibinfo {volume} {3}},\ \bibinfo {pages} {010309} (\bibinfo {year}
  {2022})}\BibitemShut {NoStop}%
\bibitem [{\citenamefont {Piveteau}\ and\ \citenamefont
  {Sutter}(2023)}]{piveteau2023circuit}%
  \BibitemOpen
  \bibfield  {author} {\bibinfo {author} {\bibfnamefont {C.}~\bibnamefont
  {Piveteau}}\ and\ \bibinfo {author} {\bibfnamefont {D.}~\bibnamefont
  {Sutter}},\ }\bibfield  {title} {\bibinfo {title} {Circuit knitting with
  classical communication},\ }\href@noop {} {\bibfield  {journal} {\bibinfo
  {journal} {IEEE Transactions on Information Theory}\ } (\bibinfo {year}
  {2023})}\BibitemShut {NoStop}%
\bibitem [{\citenamefont {Katabarwa}\ \emph {et~al.}(2024)\citenamefont
  {Katabarwa}, \citenamefont {Gratsea}, \citenamefont {Caesura},\ and\
  \citenamefont {Johnson}}]{katabarwa2024early}%
  \BibitemOpen
  \bibfield  {author} {\bibinfo {author} {\bibfnamefont {A.}~\bibnamefont
  {Katabarwa}}, \bibinfo {author} {\bibfnamefont {K.}~\bibnamefont {Gratsea}},
  \bibinfo {author} {\bibfnamefont {A.}~\bibnamefont {Caesura}},\ and\ \bibinfo
  {author} {\bibfnamefont {P.~D.}\ \bibnamefont {Johnson}},\ }\bibfield
  {title} {\bibinfo {title} {Early fault-tolerant quantum computing},\
  }\href@noop {} {\bibfield  {journal} {\bibinfo  {journal} {PRX Quantum}\
  }\textbf {\bibinfo {volume} {5}},\ \bibinfo {pages} {020101} (\bibinfo {year}
  {2024})}\BibitemShut {NoStop}%
\bibitem [{\citenamefont {Avron}\ \emph {et~al.}(2021)\citenamefont {Avron},
  \citenamefont {Casper},\ and\ \citenamefont {Rozen}}]{avron2021quantum}%
  \BibitemOpen
  \bibfield  {author} {\bibinfo {author} {\bibfnamefont {J.}~\bibnamefont
  {Avron}}, \bibinfo {author} {\bibfnamefont {O.}~\bibnamefont {Casper}},\ and\
  \bibinfo {author} {\bibfnamefont {I.}~\bibnamefont {Rozen}},\ }\bibfield
  {title} {\bibinfo {title} {Quantum advantage and noise reduction in
  distributed quantum computing},\ }\href@noop {} {\bibfield  {journal}
  {\bibinfo  {journal} {Physical Review A}\ }\textbf {\bibinfo {volume}
  {104}},\ \bibinfo {pages} {052404} (\bibinfo {year} {2021})}\BibitemShut
  {NoStop}%
\bibitem [{\citenamefont {Lee}\ and\ \citenamefont {Park}(2023)}]{Lee_2023}%
  \BibitemOpen
  \bibfield  {author} {\bibinfo {author} {\bibfnamefont {C.}~\bibnamefont
  {Lee}}\ and\ \bibinfo {author} {\bibfnamefont {D.~K.}\ \bibnamefont {Park}},\
  }\bibfield  {title} {\bibinfo {title} {Scalable quantum measurement error
  mitigation via conditional independence and transfer learning},\ }\href@noop
  {} {\bibfield  {journal} {\bibinfo  {journal} {Machine Learning: Science and
  Technology}\ }\textbf {\bibinfo {volume} {4}},\ \bibinfo {pages} {045051}
  (\bibinfo {year} {2023})}\BibitemShut {NoStop}%
\bibitem [{\citenamefont {Qiu}\ \emph {et~al.}(2022)\citenamefont {Qiu},
  \citenamefont {Luo},\ and\ \citenamefont {Xiao}}]{qiu2022distributed}%
  \BibitemOpen
  \bibfield  {author} {\bibinfo {author} {\bibfnamefont {D.}~\bibnamefont
  {Qiu}}, \bibinfo {author} {\bibfnamefont {L.}~\bibnamefont {Luo}},\ and\
  \bibinfo {author} {\bibfnamefont {L.}~\bibnamefont {Xiao}},\ }\bibfield
  {title} {\bibinfo {title} {Distributed grover's algorithm},\ }\href@noop {}
  {\bibfield  {journal} {\bibinfo  {journal} {arXiv preprint arXiv:2204.10487}\
  } (\bibinfo {year} {2022})}\BibitemShut {NoStop}%
\bibitem [{\citenamefont {Li}\ \emph {et~al.}(2017)\citenamefont {Li},
  \citenamefont {Qiu}, \citenamefont {Li}, \citenamefont {Zheng},\ and\
  \citenamefont {Rong}}]{li2017application}%
  \BibitemOpen
  \bibfield  {author} {\bibinfo {author} {\bibfnamefont {K.}~\bibnamefont
  {Li}}, \bibinfo {author} {\bibfnamefont {D.}~\bibnamefont {Qiu}}, \bibinfo
  {author} {\bibfnamefont {L.}~\bibnamefont {Li}}, \bibinfo {author}
  {\bibfnamefont {S.}~\bibnamefont {Zheng}},\ and\ \bibinfo {author}
  {\bibfnamefont {Z.}~\bibnamefont {Rong}},\ }\bibfield  {title} {\bibinfo
  {title} {Application of distributed semi-quantum computing model in phase
  estimation},\ }\href@noop {} {\bibfield  {journal} {\bibinfo  {journal}
  {Information Processing Letters}\ }\textbf {\bibinfo {volume} {120}},\
  \bibinfo {pages} {23} (\bibinfo {year} {2017})}\BibitemShut {NoStop}%
\bibitem [{\citenamefont {Yimsiriwattana}\ and\ \citenamefont
  {Lomonaco~Jr}(2004)}]{yimsiriwattana2004distributed}%
  \BibitemOpen
  \bibfield  {author} {\bibinfo {author} {\bibfnamefont {A.}~\bibnamefont
  {Yimsiriwattana}}\ and\ \bibinfo {author} {\bibfnamefont {S.~J.}\
  \bibnamefont {Lomonaco~Jr}},\ }\bibfield  {title} {\bibinfo {title}
  {Distributed quantum computing: A distributed shor algorithm},\ }in\
  \href@noop {} {\emph {\bibinfo {booktitle} {Quantum Information and
  Computation II}}},\ Vol.\ \bibinfo {volume} {5436}\ (\bibinfo {organization}
  {SPIE},\ \bibinfo {year} {2004})\ pp.\ \bibinfo {pages}
  {360--372}\BibitemShut {NoStop}%
\bibitem [{\citenamefont {Xiao}\ \emph {et~al.}(2023)\citenamefont {Xiao},
  \citenamefont {Qiu}, \citenamefont {Luo},\ and\ \citenamefont
  {Mateus}}]{xiao2023distributed}%
  \BibitemOpen
  \bibfield  {author} {\bibinfo {author} {\bibfnamefont {L.}~\bibnamefont
  {Xiao}}, \bibinfo {author} {\bibfnamefont {D.}~\bibnamefont {Qiu}}, \bibinfo
  {author} {\bibfnamefont {L.}~\bibnamefont {Luo}},\ and\ \bibinfo {author}
  {\bibfnamefont {P.}~\bibnamefont {Mateus}},\ }\bibfield  {title} {\bibinfo
  {title} {Distributed quantum-classical hybrid shor's algorithm},\ }\href@noop
  {} {\bibfield  {journal} {\bibinfo  {journal} {arXiv preprint
  arXiv:2304.12100}\ } (\bibinfo {year} {2023})}\BibitemShut {NoStop}%
\bibitem [{\citenamefont {Deng}\ \emph {et~al.}(2017)\citenamefont {Deng},
  \citenamefont {Li},\ and\ \citenamefont {Sarma}}]{deng2017quantum}%
  \BibitemOpen
  \bibfield  {author} {\bibinfo {author} {\bibfnamefont {D.-L.}\ \bibnamefont
  {Deng}}, \bibinfo {author} {\bibfnamefont {X.}~\bibnamefont {Li}},\ and\
  \bibinfo {author} {\bibfnamefont {S.~D.}\ \bibnamefont {Sarma}},\ }\bibfield
  {title} {\bibinfo {title} {Quantum entanglement in neural network states},\
  }\href@noop {} {\bibfield  {journal} {\bibinfo  {journal} {Physical Review
  X}\ }\textbf {\bibinfo {volume} {7}},\ \bibinfo {pages} {021021} (\bibinfo
  {year} {2017})}\BibitemShut {NoStop}%
\bibitem [{\citenamefont {Bowles}\ \emph {et~al.}(2023)\citenamefont {Bowles},
  \citenamefont {Wright}, \citenamefont {Farkas}, \citenamefont {Killoran},\
  and\ \citenamefont {Schuld}}]{bowles2023contextuality}%
  \BibitemOpen
  \bibfield  {author} {\bibinfo {author} {\bibfnamefont {J.}~\bibnamefont
  {Bowles}}, \bibinfo {author} {\bibfnamefont {V.~J.}\ \bibnamefont {Wright}},
  \bibinfo {author} {\bibfnamefont {M.}~\bibnamefont {Farkas}}, \bibinfo
  {author} {\bibfnamefont {N.}~\bibnamefont {Killoran}},\ and\ \bibinfo
  {author} {\bibfnamefont {M.}~\bibnamefont {Schuld}},\ }\bibfield  {title}
  {\bibinfo {title} {Contextuality and inductive bias in quantum machine
  learning},\ }\href@noop {} {\bibfield  {journal} {\bibinfo  {journal} {arXiv
  preprint arXiv:2302.01365}\ } (\bibinfo {year} {2023})}\BibitemShut {NoStop}%
\bibitem [{\citenamefont {Biamonte}\ \emph {et~al.}(2017)\citenamefont
  {Biamonte}, \citenamefont {Wittek}, \citenamefont {Pancotti}, \citenamefont
  {Rebentrost}, \citenamefont {Wiebe},\ and\ \citenamefont
  {Lloyd}}]{biamonte2017quantum}%
  \BibitemOpen
  \bibfield  {author} {\bibinfo {author} {\bibfnamefont {J.}~\bibnamefont
  {Biamonte}}, \bibinfo {author} {\bibfnamefont {P.}~\bibnamefont {Wittek}},
  \bibinfo {author} {\bibfnamefont {N.}~\bibnamefont {Pancotti}}, \bibinfo
  {author} {\bibfnamefont {P.}~\bibnamefont {Rebentrost}}, \bibinfo {author}
  {\bibfnamefont {N.}~\bibnamefont {Wiebe}},\ and\ \bibinfo {author}
  {\bibfnamefont {S.}~\bibnamefont {Lloyd}},\ }\bibfield  {title} {\bibinfo
  {title} {Quantum machine learning},\ }\href@noop {} {\bibfield  {journal}
  {\bibinfo  {journal} {Nature}\ }\textbf {\bibinfo {volume} {549}},\ \bibinfo
  {pages} {195} (\bibinfo {year} {2017})}\BibitemShut {NoStop}%
\bibitem [{\citenamefont {Havl{\'\i}{\v{c}}ek}\ \emph
  {et~al.}(2019)\citenamefont {Havl{\'\i}{\v{c}}ek}, \citenamefont
  {C{\'o}rcoles}, \citenamefont {Temme}, \citenamefont {Harrow}, \citenamefont
  {Kandala}, \citenamefont {Chow},\ and\ \citenamefont
  {Gambetta}}]{havlivcek2019supervised}%
  \BibitemOpen
  \bibfield  {author} {\bibinfo {author} {\bibfnamefont {V.}~\bibnamefont
  {Havl{\'\i}{\v{c}}ek}}, \bibinfo {author} {\bibfnamefont {A.~D.}\
  \bibnamefont {C{\'o}rcoles}}, \bibinfo {author} {\bibfnamefont
  {K.}~\bibnamefont {Temme}}, \bibinfo {author} {\bibfnamefont {A.~W.}\
  \bibnamefont {Harrow}}, \bibinfo {author} {\bibfnamefont {A.}~\bibnamefont
  {Kandala}}, \bibinfo {author} {\bibfnamefont {J.~M.}\ \bibnamefont {Chow}},\
  and\ \bibinfo {author} {\bibfnamefont {J.~M.}\ \bibnamefont {Gambetta}},\
  }\bibfield  {title} {\bibinfo {title} {Supervised learning with
  quantum-enhanced feature spaces},\ }\href@noop {} {\bibfield  {journal}
  {\bibinfo  {journal} {Nature}\ }\textbf {\bibinfo {volume} {567}},\ \bibinfo
  {pages} {209} (\bibinfo {year} {2019})}\BibitemShut {NoStop}%
\bibitem [{\citenamefont {Cerezo}\ \emph {et~al.}(2022)\citenamefont {Cerezo},
  \citenamefont {Verdon}, \citenamefont {Huang}, \citenamefont {Cincio},\ and\
  \citenamefont {Coles}}]{cerezo2022challenges}%
  \BibitemOpen
  \bibfield  {author} {\bibinfo {author} {\bibfnamefont {M.}~\bibnamefont
  {Cerezo}}, \bibinfo {author} {\bibfnamefont {G.}~\bibnamefont {Verdon}},
  \bibinfo {author} {\bibfnamefont {H.-Y.}\ \bibnamefont {Huang}}, \bibinfo
  {author} {\bibfnamefont {L.}~\bibnamefont {Cincio}},\ and\ \bibinfo {author}
  {\bibfnamefont {P.~J.}\ \bibnamefont {Coles}},\ }\bibfield  {title} {\bibinfo
  {title} {Challenges and opportunities in quantum machine learning},\
  }\href@noop {} {\bibfield  {journal} {\bibinfo  {journal} {Nature
  Computational Science}\ }\textbf {\bibinfo {volume} {2}},\ \bibinfo {pages}
  {567} (\bibinfo {year} {2022})}\BibitemShut {NoStop}%
\bibitem [{\citenamefont {Abbas}\ \emph {et~al.}(2021)\citenamefont {Abbas},
  \citenamefont {Sutter}, \citenamefont {Zoufal}, \citenamefont {Lucchi},
  \citenamefont {Figalli},\ and\ \citenamefont {Woerner}}]{abbas2021power}%
  \BibitemOpen
  \bibfield  {author} {\bibinfo {author} {\bibfnamefont {A.}~\bibnamefont
  {Abbas}}, \bibinfo {author} {\bibfnamefont {D.}~\bibnamefont {Sutter}},
  \bibinfo {author} {\bibfnamefont {C.}~\bibnamefont {Zoufal}}, \bibinfo
  {author} {\bibfnamefont {A.}~\bibnamefont {Lucchi}}, \bibinfo {author}
  {\bibfnamefont {A.}~\bibnamefont {Figalli}},\ and\ \bibinfo {author}
  {\bibfnamefont {S.}~\bibnamefont {Woerner}},\ }\bibfield  {title} {\bibinfo
  {title} {The power of quantum neural networks},\ }\href@noop {} {\bibfield
  {journal} {\bibinfo  {journal} {Nature Computational Science}\ }\textbf
  {\bibinfo {volume} {1}},\ \bibinfo {pages} {403} (\bibinfo {year}
  {2021})}\BibitemShut {NoStop}%
\bibitem [{\citenamefont {Pira}\ and\ \citenamefont
  {Ferrie}(2023)}]{pira2023invitation}%
  \BibitemOpen
  \bibfield  {author} {\bibinfo {author} {\bibfnamefont {L.}~\bibnamefont
  {Pira}}\ and\ \bibinfo {author} {\bibfnamefont {C.}~\bibnamefont {Ferrie}},\
  }\bibfield  {title} {\bibinfo {title} {An invitation to distributed quantum
  neural networks},\ }\href@noop {} {\bibfield  {journal} {\bibinfo  {journal}
  {Quantum Machine Intelligence}\ }\textbf {\bibinfo {volume} {5}},\ \bibinfo
  {pages} {1} (\bibinfo {year} {2023})}\BibitemShut {NoStop}%
\bibitem [{\citenamefont {Marshall}\ \emph {et~al.}(2023)\citenamefont
  {Marshall}, \citenamefont {Gyurik},\ and\ \citenamefont
  {Dunjko}}]{marshall2023high}%
  \BibitemOpen
  \bibfield  {author} {\bibinfo {author} {\bibfnamefont {S.~C.}\ \bibnamefont
  {Marshall}}, \bibinfo {author} {\bibfnamefont {C.}~\bibnamefont {Gyurik}},\
  and\ \bibinfo {author} {\bibfnamefont {V.}~\bibnamefont {Dunjko}},\
  }\bibfield  {title} {\bibinfo {title} {High dimensional quantum machine
  learning with small quantum computers},\ }\href@noop {} {\bibfield  {journal}
  {\bibinfo  {journal} {Quantum}\ }\textbf {\bibinfo {volume} {7}},\ \bibinfo
  {pages} {1078} (\bibinfo {year} {2023})}\BibitemShut {NoStop}%
\bibitem [{\citenamefont {Du}\ \emph {et~al.}(2021)\citenamefont {Du},
  \citenamefont {Qian},\ and\ \citenamefont {Tao}}]{du2021accelerating}%
  \BibitemOpen
  \bibfield  {author} {\bibinfo {author} {\bibfnamefont {Y.}~\bibnamefont
  {Du}}, \bibinfo {author} {\bibfnamefont {Y.}~\bibnamefont {Qian}},\ and\
  \bibinfo {author} {\bibfnamefont {D.}~\bibnamefont {Tao}},\ }\bibfield
  {title} {\bibinfo {title} {Accelerating variational quantum algorithms with
  multiple quantum processors},\ }\href@noop {} {\bibfield  {journal} {\bibinfo
   {journal} {arXiv preprint arXiv:2106.12819}\ } (\bibinfo {year}
  {2021})}\BibitemShut {NoStop}%
\bibitem [{\citenamefont {Chen}\ and\ \citenamefont
  {Yoo}(2021)}]{chen2021federated}%
  \BibitemOpen
  \bibfield  {author} {\bibinfo {author} {\bibfnamefont {S.~Y.-C.}\
  \bibnamefont {Chen}}\ and\ \bibinfo {author} {\bibfnamefont {S.}~\bibnamefont
  {Yoo}},\ }\bibfield  {title} {\bibinfo {title} {Federated quantum machine
  learning},\ }\href@noop {} {\bibfield  {journal} {\bibinfo  {journal}
  {Entropy}\ }\textbf {\bibinfo {volume} {23}},\ \bibinfo {pages} {460}
  (\bibinfo {year} {2021})}\BibitemShut {NoStop}%
\bibitem [{\citenamefont {Neumann}\ and\ \citenamefont
  {Wezeman}(2022)}]{neumann2022distributed}%
  \BibitemOpen
  \bibfield  {author} {\bibinfo {author} {\bibfnamefont {N.~M.}\ \bibnamefont
  {Neumann}}\ and\ \bibinfo {author} {\bibfnamefont {R.~S.}\ \bibnamefont
  {Wezeman}},\ }\bibfield  {title} {\bibinfo {title} {Distributed quantum
  machine learning},\ }in\ \href@noop {} {\emph {\bibinfo {booktitle}
  {International Conference on Innovations for Community Services}}}\ (\bibinfo
  {organization} {Springer},\ \bibinfo {year} {2022})\ pp.\ \bibinfo {pages}
  {281--293}\BibitemShut {NoStop}%
\bibitem [{\citenamefont {Gottesman}\ and\ \citenamefont
  {Chuang}(1999)}]{gottesman1999demonstrating}%
  \BibitemOpen
  \bibfield  {author} {\bibinfo {author} {\bibfnamefont {D.}~\bibnamefont
  {Gottesman}}\ and\ \bibinfo {author} {\bibfnamefont {I.~L.}\ \bibnamefont
  {Chuang}},\ }\bibfield  {title} {\bibinfo {title} {Demonstrating the
  viability of universal quantum computation using teleportation and
  single-qubit operations},\ }\href@noop {} {\bibfield  {journal} {\bibinfo
  {journal} {Nature}\ }\textbf {\bibinfo {volume} {402}},\ \bibinfo {pages}
  {390} (\bibinfo {year} {1999})}\BibitemShut {NoStop}%
\bibitem [{\citenamefont {Eisert}\ \emph {et~al.}(2000)\citenamefont {Eisert},
  \citenamefont {Jacobs}, \citenamefont {Papadopoulos},\ and\ \citenamefont
  {Plenio}}]{eisert2000optimal}%
  \BibitemOpen
  \bibfield  {author} {\bibinfo {author} {\bibfnamefont {J.}~\bibnamefont
  {Eisert}}, \bibinfo {author} {\bibfnamefont {K.}~\bibnamefont {Jacobs}},
  \bibinfo {author} {\bibfnamefont {P.}~\bibnamefont {Papadopoulos}},\ and\
  \bibinfo {author} {\bibfnamefont {M.~B.}\ \bibnamefont {Plenio}},\ }\bibfield
   {title} {\bibinfo {title} {Optimal local implementation of nonlocal quantum
  gates},\ }\href@noop {} {\bibfield  {journal} {\bibinfo  {journal} {Physical
  Review A}\ }\textbf {\bibinfo {volume} {62}},\ \bibinfo {pages} {052317}
  (\bibinfo {year} {2000})}\BibitemShut {NoStop}%
\bibitem [{\citenamefont {Chou}\ \emph {et~al.}(2018)\citenamefont {Chou},
  \citenamefont {Blumoff}, \citenamefont {Wang}, \citenamefont {Reinhold},
  \citenamefont {Axline}, \citenamefont {Gao}, \citenamefont {Frunzio},
  \citenamefont {Devoret}, \citenamefont {Jiang},\ and\ \citenamefont
  {Schoelkopf}}]{chou2018deterministic}%
  \BibitemOpen
  \bibfield  {author} {\bibinfo {author} {\bibfnamefont {K.~S.}\ \bibnamefont
  {Chou}}, \bibinfo {author} {\bibfnamefont {J.~Z.}\ \bibnamefont {Blumoff}},
  \bibinfo {author} {\bibfnamefont {C.~S.}\ \bibnamefont {Wang}}, \bibinfo
  {author} {\bibfnamefont {P.~C.}\ \bibnamefont {Reinhold}}, \bibinfo {author}
  {\bibfnamefont {C.~J.}\ \bibnamefont {Axline}}, \bibinfo {author}
  {\bibfnamefont {Y.~Y.}\ \bibnamefont {Gao}}, \bibinfo {author} {\bibfnamefont
  {L.}~\bibnamefont {Frunzio}}, \bibinfo {author} {\bibfnamefont
  {M.}~\bibnamefont {Devoret}}, \bibinfo {author} {\bibfnamefont
  {L.}~\bibnamefont {Jiang}},\ and\ \bibinfo {author} {\bibfnamefont
  {R.}~\bibnamefont {Schoelkopf}},\ }\bibfield  {title} {\bibinfo {title}
  {Deterministic teleportation of a quantum gate between two logical qubits},\
  }\href@noop {} {\bibfield  {journal} {\bibinfo  {journal} {Nature}\ }\textbf
  {\bibinfo {volume} {561}},\ \bibinfo {pages} {368} (\bibinfo {year}
  {2018})}\BibitemShut {NoStop}%
\bibitem [{\citenamefont {Cacciapuoti}\ \emph {et~al.}(2019)\citenamefont
  {Cacciapuoti}, \citenamefont {Caleffi}, \citenamefont {Tafuri}, \citenamefont
  {Cataliotti}, \citenamefont {Gherardini},\ and\ \citenamefont
  {Bianchi}}]{cacciapuoti2019quantum}%
  \BibitemOpen
  \bibfield  {author} {\bibinfo {author} {\bibfnamefont {A.~S.}\ \bibnamefont
  {Cacciapuoti}}, \bibinfo {author} {\bibfnamefont {M.}~\bibnamefont
  {Caleffi}}, \bibinfo {author} {\bibfnamefont {F.}~\bibnamefont {Tafuri}},
  \bibinfo {author} {\bibfnamefont {F.~S.}\ \bibnamefont {Cataliotti}},
  \bibinfo {author} {\bibfnamefont {S.}~\bibnamefont {Gherardini}},\ and\
  \bibinfo {author} {\bibfnamefont {G.}~\bibnamefont {Bianchi}},\ }\bibfield
  {title} {\bibinfo {title} {Quantum internet: Networking challenges in
  distributed quantum computing},\ }\href@noop {} {\bibfield  {journal}
  {\bibinfo  {journal} {IEEE Network}\ }\textbf {\bibinfo {volume} {34}},\
  \bibinfo {pages} {137} (\bibinfo {year} {2019})}\BibitemShut {NoStop}%
\bibitem [{\citenamefont {Zhong}\ \emph {et~al.}(2021)\citenamefont {Zhong},
  \citenamefont {Chang}, \citenamefont {Bienfait}, \citenamefont {Dumur},
  \citenamefont {Chou}, \citenamefont {Conner}, \citenamefont {Grebel},
  \citenamefont {Povey}, \citenamefont {Yan}, \citenamefont {Schuster} \emph
  {et~al.}}]{zhong2021deterministic}%
  \BibitemOpen
  \bibfield  {author} {\bibinfo {author} {\bibfnamefont {Y.}~\bibnamefont
  {Zhong}}, \bibinfo {author} {\bibfnamefont {H.-S.}\ \bibnamefont {Chang}},
  \bibinfo {author} {\bibfnamefont {A.}~\bibnamefont {Bienfait}}, \bibinfo
  {author} {\bibfnamefont {{\'E}.}~\bibnamefont {Dumur}}, \bibinfo {author}
  {\bibfnamefont {M.-H.}\ \bibnamefont {Chou}}, \bibinfo {author}
  {\bibfnamefont {C.~R.}\ \bibnamefont {Conner}}, \bibinfo {author}
  {\bibfnamefont {J.}~\bibnamefont {Grebel}}, \bibinfo {author} {\bibfnamefont
  {R.~G.}\ \bibnamefont {Povey}}, \bibinfo {author} {\bibfnamefont
  {H.}~\bibnamefont {Yan}}, \bibinfo {author} {\bibfnamefont {D.~I.}\
  \bibnamefont {Schuster}}, \emph {et~al.},\ }\bibfield  {title} {\bibinfo
  {title} {Deterministic multi-qubit entanglement in a quantum network},\
  }\href@noop {} {\bibfield  {journal} {\bibinfo  {journal} {Nature}\ }\textbf
  {\bibinfo {volume} {590}},\ \bibinfo {pages} {571} (\bibinfo {year}
  {2021})}\BibitemShut {NoStop}%
\bibitem [{\citenamefont {Lis}\ \emph {et~al.}(2023)\citenamefont {Lis},
  \citenamefont {Senoo}, \citenamefont {McGrew}, \citenamefont {R{\"o}nchen},
  \citenamefont {Jenkins},\ and\ \citenamefont {Kaufman}}]{lis2023Midcircuit}%
  \BibitemOpen
  \bibfield  {author} {\bibinfo {author} {\bibfnamefont {J.~W.}\ \bibnamefont
  {Lis}}, \bibinfo {author} {\bibfnamefont {A.}~\bibnamefont {Senoo}}, \bibinfo
  {author} {\bibfnamefont {W.~F.}\ \bibnamefont {McGrew}}, \bibinfo {author}
  {\bibfnamefont {F.}~\bibnamefont {R{\"o}nchen}}, \bibinfo {author}
  {\bibfnamefont {A.}~\bibnamefont {Jenkins}},\ and\ \bibinfo {author}
  {\bibfnamefont {A.~M.}\ \bibnamefont {Kaufman}},\ }\bibfield  {title}
  {\bibinfo {title} {Midcircuit operations using the \textit{omg} architecture
  in neutral atom arrays},\ }\href@noop {} {\bibfield  {journal} {\bibinfo
  {journal} {Phys. Rev. X.}\ }\textbf {\bibinfo {volume} {13}} (\bibinfo {year}
  {2023})}\BibitemShut {NoStop}%
\bibitem [{\citenamefont {C{\'o}rcoles}\ \emph {et~al.}(2021)\citenamefont
  {C{\'o}rcoles}, \citenamefont {Takita}, \citenamefont {Inoue}, \citenamefont
  {Lekuch}, \citenamefont {Minev}, \citenamefont {Chow},\ and\ \citenamefont
  {Gambetta}}]{corcoles2021exploiting}%
  \BibitemOpen
  \bibfield  {author} {\bibinfo {author} {\bibfnamefont {A.~D.}\ \bibnamefont
  {C{\'o}rcoles}}, \bibinfo {author} {\bibfnamefont {M.}~\bibnamefont
  {Takita}}, \bibinfo {author} {\bibfnamefont {K.}~\bibnamefont {Inoue}},
  \bibinfo {author} {\bibfnamefont {S.}~\bibnamefont {Lekuch}}, \bibinfo
  {author} {\bibfnamefont {Z.~K.}\ \bibnamefont {Minev}}, \bibinfo {author}
  {\bibfnamefont {J.~M.}\ \bibnamefont {Chow}},\ and\ \bibinfo {author}
  {\bibfnamefont {J.~M.}\ \bibnamefont {Gambetta}},\ }\bibfield  {title}
  {\bibinfo {title} {Exploiting dynamic quantum circuits in a quantum algorithm
  with superconducting qubits},\ }\href@noop {} {\bibfield  {journal} {\bibinfo
   {journal} {Physical Review Letters}\ }\textbf {\bibinfo {volume} {127}},\
  \bibinfo {pages} {100501} (\bibinfo {year} {2021})}\BibitemShut {NoStop}%
\bibitem [{\citenamefont {Haug}\ \emph {et~al.}(2021)\citenamefont {Haug},
  \citenamefont {Bharti},\ and\ \citenamefont {Kim}}]{haug2021capacity}%
  \BibitemOpen
  \bibfield  {author} {\bibinfo {author} {\bibfnamefont {T.}~\bibnamefont
  {Haug}}, \bibinfo {author} {\bibfnamefont {K.}~\bibnamefont {Bharti}},\ and\
  \bibinfo {author} {\bibfnamefont {M.}~\bibnamefont {Kim}},\ }\bibfield
  {title} {\bibinfo {title} {Capacity and quantum geometry of parametrized
  quantum circuits},\ }\href@noop {} {\bibfield  {journal} {\bibinfo  {journal}
  {PRX Quantum}\ }\textbf {\bibinfo {volume} {2}},\ \bibinfo {pages} {040309}
  (\bibinfo {year} {2021})}\BibitemShut {NoStop}%
\bibitem [{\citenamefont {Cong}\ \emph {et~al.}(2019)\citenamefont {Cong},
  \citenamefont {Choi},\ and\ \citenamefont {Lukin}}]{cong2019quantum}%
  \BibitemOpen
  \bibfield  {author} {\bibinfo {author} {\bibfnamefont {I.}~\bibnamefont
  {Cong}}, \bibinfo {author} {\bibfnamefont {S.}~\bibnamefont {Choi}},\ and\
  \bibinfo {author} {\bibfnamefont {M.~D.}\ \bibnamefont {Lukin}},\ }\bibfield
  {title} {\bibinfo {title} {Quantum convolutional neural networks},\
  }\href@noop {} {\bibfield  {journal} {\bibinfo  {journal} {Nature Physics}\
  }\textbf {\bibinfo {volume} {15}},\ \bibinfo {pages} {1273} (\bibinfo {year}
  {2019})}\BibitemShut {NoStop}%
\bibitem [{\citenamefont {Caro}\ \emph {et~al.}(2022)\citenamefont {Caro},
  \citenamefont {Huang}, \citenamefont {Cerezo}, \citenamefont {Sharma},
  \citenamefont {Sornborger}, \citenamefont {Cincio},\ and\ \citenamefont
  {Coles}}]{caro2022generalization}%
  \BibitemOpen
  \bibfield  {author} {\bibinfo {author} {\bibfnamefont {M.~C.}\ \bibnamefont
  {Caro}}, \bibinfo {author} {\bibfnamefont {H.-Y.}\ \bibnamefont {Huang}},
  \bibinfo {author} {\bibfnamefont {M.}~\bibnamefont {Cerezo}}, \bibinfo
  {author} {\bibfnamefont {K.}~\bibnamefont {Sharma}}, \bibinfo {author}
  {\bibfnamefont {A.}~\bibnamefont {Sornborger}}, \bibinfo {author}
  {\bibfnamefont {L.}~\bibnamefont {Cincio}},\ and\ \bibinfo {author}
  {\bibfnamefont {P.~J.}\ \bibnamefont {Coles}},\ }\bibfield  {title} {\bibinfo
  {title} {Generalization in quantum machine learning from few training data},\
  }\href@noop {} {\bibfield  {journal} {\bibinfo  {journal} {Nature
  communications}\ }\textbf {\bibinfo {volume} {13}},\ \bibinfo {pages} {4919}
  (\bibinfo {year} {2022})}\BibitemShut {NoStop}%
\bibitem [{\citenamefont {Holmes}\ \emph {et~al.}(2022)\citenamefont {Holmes},
  \citenamefont {Sharma}, \citenamefont {Cerezo},\ and\ \citenamefont
  {Coles}}]{holmes2022connecting}%
  \BibitemOpen
  \bibfield  {author} {\bibinfo {author} {\bibfnamefont {Z.}~\bibnamefont
  {Holmes}}, \bibinfo {author} {\bibfnamefont {K.}~\bibnamefont {Sharma}},
  \bibinfo {author} {\bibfnamefont {M.}~\bibnamefont {Cerezo}},\ and\ \bibinfo
  {author} {\bibfnamefont {P.~J.}\ \bibnamefont {Coles}},\ }\bibfield  {title}
  {\bibinfo {title} {Connecting ansatz expressibility to gradient magnitudes
  and barren plateaus},\ }\href@noop {} {\bibfield  {journal} {\bibinfo
  {journal} {PRX Quantum}\ }\textbf {\bibinfo {volume} {3}},\ \bibinfo {pages}
  {010313} (\bibinfo {year} {2022})}\BibitemShut {NoStop}%
\bibitem [{\citenamefont {Pesah}\ \emph {et~al.}(2021)\citenamefont {Pesah},
  \citenamefont {Cerezo}, \citenamefont {Wang}, \citenamefont {Volkoff},
  \citenamefont {Sornborger},\ and\ \citenamefont {Coles}}]{pesah2021absence}%
  \BibitemOpen
  \bibfield  {author} {\bibinfo {author} {\bibfnamefont {A.}~\bibnamefont
  {Pesah}}, \bibinfo {author} {\bibfnamefont {M.}~\bibnamefont {Cerezo}},
  \bibinfo {author} {\bibfnamefont {S.}~\bibnamefont {Wang}}, \bibinfo {author}
  {\bibfnamefont {T.}~\bibnamefont {Volkoff}}, \bibinfo {author} {\bibfnamefont
  {A.~T.}\ \bibnamefont {Sornborger}},\ and\ \bibinfo {author} {\bibfnamefont
  {P.~J.}\ \bibnamefont {Coles}},\ }\bibfield  {title} {\bibinfo {title}
  {Absence of barren plateaus in quantum convolutional neural networks},\
  }\href@noop {} {\bibfield  {journal} {\bibinfo  {journal} {Physical Review
  X}\ }\textbf {\bibinfo {volume} {11}},\ \bibinfo {pages} {041011} (\bibinfo
  {year} {2021})}\BibitemShut {NoStop}%
\bibitem [{\citenamefont {T{\"u}ys{\"u}z}\ \emph {et~al.}(2023)\citenamefont
  {T{\"u}ys{\"u}z}, \citenamefont {Clemente}, \citenamefont {Crippa},
  \citenamefont {Hartung}, \citenamefont {K{\"u}hn},\ and\ \citenamefont
  {Jansen}}]{tuysuz2023classical}%
  \BibitemOpen
  \bibfield  {author} {\bibinfo {author} {\bibfnamefont {C.}~\bibnamefont
  {T{\"u}ys{\"u}z}}, \bibinfo {author} {\bibfnamefont {G.}~\bibnamefont
  {Clemente}}, \bibinfo {author} {\bibfnamefont {A.}~\bibnamefont {Crippa}},
  \bibinfo {author} {\bibfnamefont {T.}~\bibnamefont {Hartung}}, \bibinfo
  {author} {\bibfnamefont {S.}~\bibnamefont {K{\"u}hn}},\ and\ \bibinfo
  {author} {\bibfnamefont {K.}~\bibnamefont {Jansen}},\ }\bibfield  {title}
  {\bibinfo {title} {Classical splitting of parametrized quantum circuits},\
  }\href@noop {} {\bibfield  {journal} {\bibinfo  {journal} {Quantum Machine
  Intelligence}\ }\textbf {\bibinfo {volume} {5}},\ \bibinfo {pages} {34}
  (\bibinfo {year} {2023})}\BibitemShut {NoStop}%
\bibitem [{\citenamefont {Bergholm}\ \emph {et~al.}(2018)\citenamefont
  {Bergholm}, \citenamefont {Izaac}, \citenamefont {Schuld}, \citenamefont
  {Gogolin}, \citenamefont {Ahmed}, \citenamefont {Ajith}, \citenamefont
  {Alam}, \citenamefont {Alonso-Linaje}, \citenamefont {AkashNarayanan},
  \citenamefont {Asadi} \emph {et~al.}}]{bergholm2018pennylane}%
  \BibitemOpen
  \bibfield  {author} {\bibinfo {author} {\bibfnamefont {V.}~\bibnamefont
  {Bergholm}}, \bibinfo {author} {\bibfnamefont {J.}~\bibnamefont {Izaac}},
  \bibinfo {author} {\bibfnamefont {M.}~\bibnamefont {Schuld}}, \bibinfo
  {author} {\bibfnamefont {C.}~\bibnamefont {Gogolin}}, \bibinfo {author}
  {\bibfnamefont {S.}~\bibnamefont {Ahmed}}, \bibinfo {author} {\bibfnamefont
  {V.}~\bibnamefont {Ajith}}, \bibinfo {author} {\bibfnamefont {M.~S.}\
  \bibnamefont {Alam}}, \bibinfo {author} {\bibfnamefont {G.}~\bibnamefont
  {Alonso-Linaje}}, \bibinfo {author} {\bibfnamefont {B.}~\bibnamefont
  {AkashNarayanan}}, \bibinfo {author} {\bibfnamefont {A.}~\bibnamefont
  {Asadi}}, \emph {et~al.},\ }\bibfield  {title} {\bibinfo {title} {Pennylane:
  Automatic differentiation of hybrid quantum-classical computations},\
  }\href@noop {} {\bibfield  {journal} {\bibinfo  {journal} {arXiv preprint
  arXiv:1811.04968}\ } (\bibinfo {year} {2018})}\BibitemShut {NoStop}%
\bibitem [{\citenamefont {Hur}\ \emph {et~al.}(2022)\citenamefont {Hur},
  \citenamefont {Kim},\ and\ \citenamefont {Park}}]{hur2022quantum}%
  \BibitemOpen
  \bibfield  {author} {\bibinfo {author} {\bibfnamefont {T.}~\bibnamefont
  {Hur}}, \bibinfo {author} {\bibfnamefont {L.}~\bibnamefont {Kim}},\ and\
  \bibinfo {author} {\bibfnamefont {D.~K.}\ \bibnamefont {Park}},\ }\bibfield
  {title} {\bibinfo {title} {Quantum convolutional neural network for classical
  data classification},\ }\href@noop {} {\bibfield  {journal} {\bibinfo
  {journal} {Quantum Machine Intelligence}\ }\textbf {\bibinfo {volume} {4}},\
  \bibinfo {pages} {3} (\bibinfo {year} {2022})}\BibitemShut {NoStop}%
\bibitem [{\citenamefont {Sim}\ \emph {et~al.}(2019)\citenamefont {Sim},
  \citenamefont {Johnson},\ and\ \citenamefont
  {Aspuru-Guzik}}]{sim2019expressibility}%
  \BibitemOpen
  \bibfield  {author} {\bibinfo {author} {\bibfnamefont {S.}~\bibnamefont
  {Sim}}, \bibinfo {author} {\bibfnamefont {P.~D.}\ \bibnamefont {Johnson}},\
  and\ \bibinfo {author} {\bibfnamefont {A.}~\bibnamefont {Aspuru-Guzik}},\
  }\bibfield  {title} {\bibinfo {title} {Expressibility and entangling
  capability of parameterized quantum circuits for hybrid quantum-classical
  algorithms},\ }\href@noop {} {\bibfield  {journal} {\bibinfo  {journal}
  {Advanced Quantum Technologies}\ }\textbf {\bibinfo {volume} {2}},\ \bibinfo
  {pages} {1900070} (\bibinfo {year} {2019})}\BibitemShut {NoStop}%
\bibitem [{\citenamefont {Huembeli}\ and\ \citenamefont
  {Dauphin}(2021)}]{huembeli2021characterizing}%
  \BibitemOpen
  \bibfield  {author} {\bibinfo {author} {\bibfnamefont {P.}~\bibnamefont
  {Huembeli}}\ and\ \bibinfo {author} {\bibfnamefont {A.}~\bibnamefont
  {Dauphin}},\ }\bibfield  {title} {\bibinfo {title} {Characterizing the loss
  landscape of variational quantum circuits},\ }\href@noop {} {\bibfield
  {journal} {\bibinfo  {journal} {Quantum Science and Technology}\ }\textbf
  {\bibinfo {volume} {6}},\ \bibinfo {pages} {025011} (\bibinfo {year}
  {2021})}\BibitemShut {NoStop}%
\bibitem [{\citenamefont {Kingma}\ and\ \citenamefont
  {Ba}(2014)}]{kingma2014adam}%
  \BibitemOpen
  \bibfield  {author} {\bibinfo {author} {\bibfnamefont {D.~P.}\ \bibnamefont
  {Kingma}}\ and\ \bibinfo {author} {\bibfnamefont {J.}~\bibnamefont {Ba}},\
  }\bibfield  {title} {\bibinfo {title} {Adam: A method for stochastic
  optimization},\ }\href@noop {} {\bibfield  {journal} {\bibinfo  {journal}
  {arXiv preprint arXiv:1412.6980}\ } (\bibinfo {year} {2014})}\BibitemShut
  {NoStop}%
\end{thebibliography}%

\end{document}